\documentclass{JHEP3}
\usepackage{amssymb}
\usepackage{amsfonts}
\usepackage{mathrsfs}

\newcommand{\e}{\begin{eqnarray}}
\newcommand{\ee}{\end{eqnarray}}

\newcommand{\CN}{{\cal N}}
\newcommand{\CL}{{\cal L}}

\def\a{\alpha}
\def\b{\beta}
\def\d{\delta}
\newcommand{\ep}{\epsilon}
\newcommand{\g}{\gamma}
\newcommand{\p}{\psi}

\def\t{\tau}
\newcommand{\vp}{\varepsilon}

\newcommand{\cj}{\mathcal{J}}

\title{$OSp(5|4)$ Superconformal Symmetry of $\CN=5$ Chern-Simons Theory}

\author{Fa-Min Chen \\
Department of Physics, Beijing Jiaotong Univeristy, Beijing 100044, China\\
E-mail:
\email{fmchen@bjtu.edu.cn}}

\abstract{We demonstrate that the general $D=3, {\cal N}=5$ Chern-Simons matter theory possesses a full $OSp(5|4)$ superconformal symmetry, and construct the corresponding superconformal currents.  The closure of the superconformal algebra is verified in detail.  We also show that the conserved $OSp(6|4)$ superconformal currents in the general $\CN=6$ theory can be obtained as special cases of the $OSp(5|4)$ currents by enhancing the R-symmetry of the $\CN=5$ theory from $USp(4)$ to $SU(4)$.}


\usepackage{bbm}
\usepackage{amsmath}
\usepackage{slashed}

\begin{document}

\section{Introduction} \label{Introduction}

The general $\CN=5$ Chern-Simons matter (CSM) theory, as the dual gauge description of multi M2-branes was first constructed in Ref. \cite{Hosomichi:2008jb}. It was also showed that the same physical theory can be built up in terms of the symplectic 3-algebra \cite{Chen2,ChenWu3}.

According to the gauge/M-theory dual, we expect that the $\CN=5$ theory is not only super poincare invariant but also superconformal invariant\footnote{In this paper, the super poincare transformations will be denoted as $\d_\ep$, satisfying $[\d_{\ep_1},\d_{\ep_2}]\sim P_\mu$, with $P_\mu$ the translations; and the superconformal transformations will be denoted as $\d_\eta$, satisfying $[\d_{\eta_1},\d_{\eta_2}]\sim K_\mu$, with $K_\mu$ the special conformal transformations. The full super transformations (containing both $\d_\ep$ \emph{and} $\d_\eta$) will be called the $OSp(5|4)$ superconformal transformations.}, i.e. it should have a full $OSp(5|4)$ superconformal symmetry, just as that the $\CN=6$ ABJM theory \cite{ABJM} has a complete $OSp(6|4)$ symmetry \cite{Schwarz}. But to our knowledge, only the poincare supersymmetry transformations of the $\CN=5$ theory are given (see \cite{Hosomichi:2008jb,Chen2,ChenWu3}) in the literature. The action of the $\CN=5$ theory is also invariant under scale transformations. However,  classical scale invariance plus  super-poincare invariance does \emph{not} necessarily imply superconformal invariance. It is therefore important to derive the law of superconformal transformations of the $\CN=5$ theory, and to verify that the action is invariant under the transformations. This is completed in Section \ref{SecN5}. In addition, we derive the explicit expressions of the $OSp(5|4)$ superconformal currents. These currents may be useful in that one can use them to construct the $\CN=5$ M2-brane superconformal algebras, and to study the BPS brane configurations.

We also verify the closure of the $OSp(5|4)$ superconformal algebra. In the literature, only the closure of the $\CN=5$ poincare algebra was verified explicitly \cite{Chen2}. Thus our verification the $OSp(5|4)$ superconformal algebra will fill this gap. The law of transformations generated by the bosonic algebra of $OSp(5|4)$ bosonic transformations can be read off from our calculation of the commutators of two $OSp(5|4)$ superconformal transformations.

On the other hand, since the $\CN=6$ theory is a special case of the $\CN=5$ theory, one should be able to derive the $OSp(6|4)$ superconformal currents from that of $\CN=5$. This is indeed the case: We demonstrate that the $OSp(6|4)$ superconformal currents can be obtained as special cases of the $OSp(5|4)$ ones by enhancing the supersymmetry from $\CN=5$ to $\CN=6$. In particular, we show that if we choose $U(N)\times U(N)$ as the gauge group, these $OSp(6|4)$ superconformal currents are in agreement with the previous results derived directly from the ABJM theory \cite{Schwarz}, as expected.

The $OSp(8|4)$ superconformal currents associated with the $\CN=8$ BLG theory \cite{Bagger,Gustavsson} were first constructed in Ref. \cite{Schwarz0}. We argue that they can be re-derived as special examples of the $OSp(6|4)$ currents associated with the general $\CN=6$ theory, since the $\CN=6$ supersymmetry can be promoted to $\CN=8$ if one chooses the bosonic subalgbra of $PSU(2|2)$, i.e. $SU(2)\times SU(2)$, as the Lie algebra of the gauge symmetry \cite{ABJM}.

The paper is organized as follows. In Section \ref{SecN5},
we verify that the general $\CN=5$ theory has the full $OSp(5|4)$
 superconformal symmetry and construct the corresponding conserved supercurrents.
In Section \ref{SecN6}, we derive the $OSp(6|4)$ superconformal currents by enhancing the R-symmetry from $SO(5)$ to $SO(6)$. In Section \ref{SecClosure}, the closure of the $OSp(5|4)$ superconformal algebra is verified. Section \ref{SecDis} is devoted to conclusions. Our conventions and some useful identities can be found in Appendix \ref{Identities}. In Appendix \ref{SecN5Action}, we review the general $\CN=5$ theory. In Appendix \ref{SecDN5C}, we present the details of the derivation of the $\CN=5$ super poincare currents associated with the $\CN=5$ theory.

\section{$OSp(5|4)$ Superconformal Currents}\label{SecN5}
\subsection{General Construction}\label{SecN51}
In this section,
we prove that the general $\CN=5$ theory has a full $OSp(5|4)$ symmetry, and derive the conserved $OSp(5|4)$ superconformal currents. We begin by considering the poincare supersymmetry transformation of the action (see Appendix \ref{SecN5Action} for a review of the $\CN=5$ theory). We denote the set of parameters of the $\CN=5$ poincare supersymmetry transformations as $\ep^I$ ($I=1,\ldots, 5$). According to the standard Noether method, if one allows $\ep^I$ to depend on the spacetime coordinates, the super variation of the action (\ref{5lagran}) must take the form
\e\label{EdS}
\d_\ep S=\int d^3x(-j^I_\mu)\partial^\mu\ep^I,
\ee
since if the action is invariant under the $\CN=5$ poincare supersymmetry transformations (\ref{n5susy}), (\ref{EdS}) must vanish when $\ep^I$ are constants. If the equations of motion are obeyed, the right hand side of (\ref{EdS}) must vanish; integrating by parts, we obtain the conserved currents $\partial^\mu j^I_\mu=0.$ The details of the calculation of (\ref{EdS}) will be presented in Appendix \ref{SecDN5C}.
Writing the super variation of fermionic fields (see (\ref{n5susy})) as $\d_\ep\psi^{a}_A= (\d\psi)^{Ia}_A\ep^I$, where
\begin{equation}\label{svofp}
(\d\psi)^{Ia}_A\equiv-\gamma^{\mu}D_\mu Z^a_B\Gamma^I_A{}^B
-\frac{1}{3}k_{mn}\t^{ma}{}_{b}\omega^{BC}Z^b_B\mu^n_{CD}\Gamma^I_A{}^D
+\frac{2}{3}k_{mn}\t^{ma}{}_{b}\omega^{BD}Z^b_C\mu^n_{DA}\Gamma^I_B{}^C,
\end{equation}
(The quantities appeare in (\ref{svofp}) are defined in Appendix \ref{SecN5Action}.) then the $\CN=5$ poincare supercurrents are given by
\e\label{superpcu}
j^I_\mu=-i\bar\psi^A_a\g_\mu(\d\psi)^{Ia}_A.
\ee

As usual, the set of conserved charges are defined as
\e\label{cscha}
Q^I=-\int d^2xj^{I}_0.
\ee
If we impose the equal-time commutators
\e\label{fmcommu}
&&\{\bar\psi^A_a(t,\vec{x}^\prime),\psi^b_B(t,\vec{x})\}
=-\d^A_B\d^b_a\gamma^0\d^2(x-x^\prime),\\
&&[\Pi^A_a(t,\vec{x}^\prime),Z^b_B(t,\vec{x})]=-i\d^A_B\d^b_a\d^2(x-x^\prime)
,\\
&&[\Pi^\mu_m(t,\vec{x}^\prime),A^n_\nu(t,\vec{x})]=-i\d^n_m\d^\mu_\nu\d^2(x-x^\prime),
\ee
where $\Pi^A_a(t,\vec{x}^\prime)=D_0\bar Z^A_a(t,\vec{x}^\prime)$ and $\Pi^\mu_m(t,\vec{x}^\prime)=\ep^{\lambda 0\mu}k_{mp}A^p_\lambda(t,\vec{x}^\prime)$,
then the variation of an arbitrary field $\Phi$ can be defined as
\e\label{sptrans}
\d_\ep\Phi=[-i\ep^IQ^I,\Phi].
\ee
Here $\Phi$ can be a scalar, fermion, or gauge field. It is not difficult to show
that (\ref{sptrans}) gives the law of super poincare transformations (\ref{n5susy})
by using the canonical commutators. For instance, if $\Phi=\psi^b_B(t,\vec{x})$, using the canonical anti-commutator (\ref{fmcommu}), one can easily prove that (\ref{sptrans}) gives the law of the supersymmetry transformations of the fermionic fields (see the second equation of (\ref{n5susy})).

Without changing the physical content, it is possible to introduce the modified currents
\e\label{pjcu}
\tilde{j}^I_\mu=j^I_\mu-\frac{i}{4}\partial^\nu(Z^a_A\bar\psi^B_a\Gamma^I_B{}^A)[\g_\mu, \g_\nu].
\ee
The charges remain the same, i.e. $Q^I=-\int d^2x\tilde{j}^I_0=-\int d^2xj^I_0$, since the total derivative term in (\ref{pjcu}) does not contribute to the integral. On the other hand, the modified currents $\tilde{j}^I_\mu$ are also conserved, i.e.  $\partial^\mu \tilde j^I_\mu=0$, on account of that $[\g_\mu, \g_\nu]$ is antisymmetric in $\mu$ and $\nu$.  However, what interests us most is that $\tilde{j}^I_\mu$ are  $\g$-traceless
\e\label{GTraceless}
\tilde{j}^I_\mu\g^\mu=0,
\ee
provided that the equations of motion of the fermionic fields are satisfied \footnote{We only require that the currents $\tilde{j}^I_\mu$ are  conserved on-shell.}. This allows us to define the new currents
\e\label{scu}
\tilde{s}^I_\mu=-\tilde{j}^I_\mu\g\cdot x.
\ee
Using (\ref{GTraceless}) and $\partial^\mu \tilde j^I_\mu=0$, one can immediately prove that $\tilde{s}^I_\mu$ are conserved: $\partial^\mu \tilde{s}^I_\mu=0$.


Dimensional analysis suggests that the
conserved charges
\e\label{eq}
S^I=-\int d^2x\tilde{s}^I_0
\ee
generate the superconformal transformations. Indeed, in Section \ref{SecClosure}, our calculation shows that $[\d_{\eta_1}, \d_{\eta_2}]$ does generate the special conformal transformation $K_\mu$, where $\d_{\eta_1}$ is the transformation generated by (\ref{eq}).

We are now ready to derive the superconformal transformations of the matter fields and gauge fields. Plugging (\ref{pjcu}) into (\ref{scu}), we note that (\ref{scu}) can be rewritten as
\e\label{scu2}
\tilde{s}^I_\mu=- j^I_\mu\g\cdot x -iZ^a_B\bar\psi^A_a\g_\mu\Gamma^I_A{}^B+\partial^\nu\big(\frac{i}{4}Z^a_A\bar\psi^B_a\Gamma^I_B{}^A[\g_\mu, \g_\nu]x\cdot\gamma\big).
\ee
Notice that the last term is just a total derivative term. As a result, we have the conserved charges
\e\label{scha}
S^I=-\int d^2x\tilde{s}^I_0=\int d^2x\big( j^I_0\g\cdot x+iZ^a_B\bar\psi^A_a\g_{0}\Gamma^I_A{}^B\big).
\ee
In analogue to (\ref{sptrans}), we define the superconformal variation an arbitrary field $\Phi$ as follows
\e\label{scftrans}
\d_\eta\Phi=[-i\eta^IS^I,\Phi],
\ee
where $\eta^I$ are a set of parameters, with $I=1,\ldots,5$ a fundamental index of $SO(5)$. Using (\ref{cscha})$-$(\ref{sptrans}), (\ref{scha}), and (\ref{n5susy}), one can readily derive the $\CN=5$ superconformal transformations
\begin{eqnarray}\label{n5csusy}\nonumber
\delta_\eta Z^a_A&=&i\g\cdot x\eta_A{}^{B}\psi^a_{B},\nonumber\\
\delta_\eta\psi^a_{A}&=&(\d\psi)^{Ia}_{A}\g\cdot x\eta^I-\eta_A{}^BZ^a_B,
\\
\delta_\eta A^m_\mu&=& i(\g\cdot x\eta^{AB})\gamma_\mu \cj^m_{AB},\nonumber
\end{eqnarray}
where $(\d\psi)^{Ia}_{A}$ is defined by (\ref{svofp}), and the set of antisymmetric parameters $\eta_A{}^B=\eta^I\Gamma^I_A{}^B$ ($I=1,\ldots, 5$) satisfy the traceless conditions and the reality conditions
\begin{eqnarray}\label{CSusyPara}
\omega_{AB}\eta^{AB}=0 ,\quad
\eta^{*}_{AB}=\omega^{AC}\omega^{BD}\eta_{CD}.
\end{eqnarray}

Alternatively, we can derive the $\CN=5$ superconformal transformations (\ref{n5csusy})  and currents (\ref{scu}) by adopting the method used to construct the $\CN=6$ superconformal transformations and the corresponding currents of the ABJM theory \cite{Schwarz}. Replacing $\ep^I$ in the variation of the action (\ref{EdS}) by $\g\cdot x\eta^I$,
\e
\ep^I\rightarrow \g\cdot x\eta^I,
\ee
with $\eta^I$ \emph{independent} of $x^\mu$, a short calculation shows that (\ref{EdS}) becomes
\e\label{remain}
-\int d^2x [(j^I_\mu\g^\mu )\eta^I],
\ee
where $j^I_\mu$ are the super poincare currents (\ref{superpcu}). The remaining term (\ref{remain}) can be canceled by adding
\e\label{atranpsi}
\d^\prime_{\eta}\psi^a_A=-\eta_A{}^BZ^a_B
\ee
into the transformation of the fermion field. In other words, the action is also invariant under the new transformations, defined by replacing $\ep^I$ by $\g\cdot x\eta^I$ in the $\CN=5$ super poincare transformations (\ref{n5susy}) and adding the additional term (\ref{atranpsi}) into the transformation of the fermion field. The new transformations defined in this way are nothing but the $\CN=5$ superconformal transformations (\ref{n5csusy}). And the corresponding currents, derived by using the trick $\ep^I\rightarrow \g\cdot x\eta^I$, are given by
\e\label{scu3}
s^I_\mu=-j^I_\mu\g\cdot x -iZ^a_B\bar\psi^A_a\g_\mu\Gamma^I_A{}^B,
\ee
which are exactly the same as the first two terms of $\tilde{s}^I_\mu$ defined by (\ref{scu2}). However, since the last term of (\ref{scu2}) is a conserved total derivative term \footnote{Witting the last term of (\ref{scu2}) as $\partial^\nu (A_{\mu\nu})$, we have $\partial^\mu\partial^\nu (A_{\mu\nu})=0$ immediately, due to the fact that $A_{\mu\nu}=-A_{\nu\mu}$.}, we are led to $\partial^\mu s^I_\mu=\partial^\mu\tilde s^I_\mu=0$ and $S^I=-\int d^2xs^I_0=-\int d^2x\tilde s^I_0$. Hence the conserved currents $s^I_\mu$ are \emph{equivalent} to $\tilde{s}^I_\mu$ defined by (\ref{scu2}) or (\ref{scu}).

\subsection{Currents in the $\CN=5$, $USp(2N)\times SO(M)$ Theory}\label{secN52}
If one specifies the superalgebra (\ref{SLie}) as $OSp(M|2N)$, or in other words, if one chooses the bosonic subalgebra of $OSp(M|2N)$ as the Lie algebra of the gauge symmetry, then the gauge group of the $\CN=5$ theory in Appendix \ref{SecN5Action} becomes $USp(2N)\times SO(M)$ \cite{Hosomichi:2008jb}. The same theory was constructed in a 3-algebra framework \cite{Chen2,Palmkvist2}. The $\CN=5$ theory with $USp(2N)\times SO(M)$ gauge group was
conjectured to be the gauge description of multi M2-branes in orbifold
$\textbf{C}^4/\hat{\textbf{D}}_k$, where
$\hat{\textbf{D}}_k$ is the binary dihedral group
\cite{Hosomichi:2008jb, Aharony:2008gk}; the dual gravity theory was investigated in Ref. \cite{Aharony:2008gk}.

The details of the construction of the $\CN=5$ theory with $USp(2N)\times SO(M)$ gauge group can be found in Ref. \cite{Hosomichi:2008jb,Chen2}. In this theory, both the scalar field $Z^{k\hat{k}}_A$ and spinor field $\psi^{k\hat{k}}_A$ take values in the bifundamental representation of $USp(2N)\times SO(M)$. (Their complex conjugates are denoted as $Z_{\hat{k}k}^{\dag A}$ and $\psi_{\hat{k}k}^{\dag A}$.) Here
$k=1,\ldots,M$ are the $SO(M)$ indices, and
$\hat{k}=1,\ldots,2N$ the $USp(2N)$ indices. The $USp(2N)$ gauge potential is denoted as $\hat{A}_\mu$, while the $SO(M)$
gauge potential is denoted as $A_\mu$.
In matrix notation, the law of $\CN=5$ super poincare transformations (\ref{n5susy}) now reads \cite{Hosomichi:2008jb,Chen2}
\begin{eqnarray}\label{Sp2NOMSusyTrans}\nonumber
\delta_\ep Z_A&=&i\epsilon_A{}^B\psi_B\nonumber\\
\delta_\ep \psi_A&=&-\gamma^{\mu}D_\mu
Z_B\epsilon_A{}^B+\frac{2k}{3}\epsilon_
A{}^C(Z_{[B}Z^{\dag
B}Z_{C]}+Z_BZ^\dag_CZ^B)\nonumber\\&&-\frac{4k}{3}\epsilon_B{}^C(Z_{[C}Z^{\dag
B}Z_{A]}+Z_CZ^\dag_AZ^B)\nonumber\\
\delta_\ep  A_\mu{}&=&ik\epsilon^{AB}\gamma_\mu
(Z_A\psi^\dag_B+\psi_BZ^\dag_A)\nonumber\\ \delta_\ep
\hat{A}_\mu{}&=&-ik\epsilon^{AB}\gamma_\mu
(\psi^\dag_BZ_A+Z^\dag_A\psi_B),
\end{eqnarray}
where $k$ is a constant.

From the second equation of (\ref{Sp2NOMSusyTrans}), we can immediately read off $(\d\psi)^{I}_A$ defined by (\ref{svofp}):
\e
(\d\psi)^{I}_A&=&-\gamma^{\mu}D_\mu
Z_B\Gamma^I_A{}^B+\frac{2k}{3}\Gamma^I_
A{}^C(Z_{[B}Z^{\dag
B}Z_{C]}+Z_BZ^\dag_CZ^B)\nonumber\\&&-\frac{4k}{3}\Gamma^I_B{}^C(Z_{[C}Z^{\dag
B}Z_{A]}+Z_CZ^\dag_AZ^B).
\ee
And the superconformal transformations (\ref{n5csusy}) now become
\begin{eqnarray}\label{n5scontran2}\nonumber
\delta_\eta Z_A&=&i\g\cdot x\eta_A{}^B\psi_B\nonumber\\
\delta_\eta \psi_A&=&(\d\psi)^{I}_A\g\cdot x\eta_A{}^B-\eta_A{}^BZ_B\nonumber\\
\delta_\eta  A_\mu{}&=&ik(\g\cdot x\eta^{AB})\gamma_\mu
(Z_A\psi^\dag_B+\psi_BZ^\dag_A)\nonumber\\
\delta_\eta
\hat{A}_\mu{}&=&-ik(\g\cdot x\eta^{AB})\gamma_\mu
(\psi^\dag_BZ_A+Z^\dag_A\psi_B).
\end{eqnarray}
The supercurrents (\ref{pjcu}) and (\ref{scu}) now become
\e\label{pjcu2}
\tilde{j}^I_\mu&=&-i{\rm Tr}(\psi^{\dag A}\g_\mu(\d\psi)^{I}_A)-\frac{i}{4}\partial^\nu{\rm Tr}(\psi^{\dag B}Z_A)\Gamma^I_B{}^A[\g_\mu, \g_\nu],\\
\tilde{s}^I_\mu&=&-\tilde{j}^I_\mu\g\cdot x,\label{sjcu2}
\ee
where the trace ``${\rm Tr}$" is defined as  ${\rm Tr}(\psi^{\dag B}Z_A)=\psi^{\dag B}_{\hat kk}Z^{k\hat k}_A$.

\section{$OSp(6|4)$ Superconformal Currents}\label{SecN6}
In this section, we will first derive the  $OSp(6|4)$ superconformal currents associated with the general $\CN=6$ theory from the $OSp(5|4)$ currents by enhancing the supersymmetry of the $\CN=5$ theory to $\CN=6$. As a check, we then will construct the $OSp(6|4)$ superconformal currents of the ABJM theory \cite{ABJM} by specifying the gauge group as the bosonic part of $U(N|N)$, and show that they are the same as the ones \cite{Schwarz} constructed directly from the ABJM theory.

\subsection{General Construction}

In Ref. \cite{Hosomichi:2008jb}, it has been shown that if the pseudo-real representation of the bosonic subalgebra of
(\ref{SLie}) can be decomposed into ($R\oplus\bar R$), i.e.  a direct sum of complex representation $R$ and its conjugate representation $\bar R$, then the $USp(4)$ R-symmetry of the $\CN=5$ theory will be enhanced to $SU(4)$. As a result, the $\CN=5$ supersymmetry will be promoted to $\CN=6$. (For a 3-algebra approach, see \cite{Chen2,Palmkvist1}.)

Specifically, the authors of Ref. \cite{Hosomichi:2008jb} have been able to derive the $\CN=6$ theory from the $\CN=5$ theory by decomposing the following $\CN=5$ quantities into $\CN=6$ quantities \footnote{To avoid introducing too many indices, we still use $a=1,\ldots, L$ to label an index of $\CN=6$ gauge group, (In the $\CN=5$ formulas, $a=1,\ldots, 2L$.), and use $A=1,\ldots, 4$ to label an $SU(4)$ index of R-symmetry in the right hand sides (RHS) of (\ref{n6fields}). We hope this will cause any confusion.
}:
\begin{equation}\label{n6fields}
(Z^a_A)_{\CN=5}=\begin{pmatrix} \bar Z^a_A  \\
\omega_{AB}Z^B_a
\end{pmatrix},\quad
(\psi^a_A)_{\CN=5}=\begin{pmatrix} \omega_{AB}\psi^{Ba}  \\
-\bar\psi_{Aa}
\end{pmatrix},
\end{equation}
and
\begin{equation}\label{n6matrices}
(\t^{ma}{}_{b})_{\CN=5}=\begin{pmatrix} \t^{ma}{}_{b} & 0\\
0 & -\t^{mb}{}_{a}
\end{pmatrix}
,\quad
(\omega_{ab})_{\CN=5}=\begin{pmatrix} 0 & \d_a{}^b \\
-\d^a{}_b & 0
\end{pmatrix}.
\end{equation}
The matrix $\t^{ma}{}_{b}$ obeys the reality condition $\t^{* ma}{}_{b}=-\t^{mb}{}_a$. The reality conditions of the $\CN=5$ theory (\ref{RealCondi}) are decomposed into
\begin{equation}\label{N6RealCondi}
Z^{*A}_a= \bar Z_A^{a},\quad \psi^{*Aa}=\bar\psi_{Aa},
\end{equation}
which are in agreement with the expected $\CN=6$ ones. In the $\CN=6$ notation, the ``fundamental identity" (\ref{EFI}) reduces to \cite{Hosomichi:2008jb}
\e\label{EFI2}
k_{mn}\t^{ma}{}_{[b}\t^{nc}{}_{d]}=0.
\ee

Using the above decompositions and the identity (\ref{totanti}), the $OSp(5|4)$ superconformal currents (\ref{pjcu}) and (\ref{scu}) \footnote{In Section  \ref{SecN5}, in particular in Eqs. (\ref{pjcu}) and (\ref{scu}), we use $I=1,\ldots,5$ to label the $SO(5)$ indices, while in this section, we use $I^\prime=1,\ldots,5$ to label the $SO(5)$ indices. However, in this section, we use $I=1,\ldots, 6$ to label the $SO(6)$ indices. We hope this will not cause any confusion.} can be expressed in terms of the $\CN=6$ fields as follows
\e\label{n6psc}
(\tilde j^{I^\prime}_\mu)_{\CN=5}&=&-i\bar\psi_{Aa}\g_\mu(\d\psi)^{I^\prime Aa}-i\psi^{Aa}\g_\mu(\d\bar\psi)^{I^\prime}_{Aa}
\nonumber\\&&
-\frac{i}{4}[\g_\mu,\g_\nu]\partial^\nu\big(\bar Z^a_A\bar\psi_{Ba}\Gamma^{I^\prime AB}+  Z^A_a\psi^{Ba}\Gamma^{I^\prime}_{AB}\big),
\\ (\tilde{s}^{I^\prime}_\mu)_{\CN=5}&=&\g\cdot x(\tilde{j}^{I^\prime}_\mu)_{\CN=5},\label{n6sccur}
\ee
where
\begin{equation}
(\d\bar\psi)^{I^\prime}_{Aa} = \gamma^\mu D_\mu Z^B_a\Gamma^{I^\prime}_{BA} +k_{mn}
  M^{mB}{}_C\t^{nb}{}_aZ^C_b\Gamma^{I^\prime}_{BA}+k_{mn}M^{mD}{}_A\t^{mb}{}_a
  Z^C_b \Gamma^{I^\prime}_{CD},
\end{equation}
and $(\d\psi)^{I^\prime Aa}$ is the complex conjugate of $(\d\bar\psi)^{I^\prime}_{Aa}$. \emph{Here we have used $I^\prime=1,\ldots,5$ to label the fundamental index of $SO(5)$.} And we have defined the $\CN=6$ ``momentum map" and
``current" operators as \cite{Hosomichi:2008jb}
\e M^{mA}{}_B\equiv\t^{ma}{}_bZ_a^A\bar{Z}^b_B,\quad
J^{mAB}\equiv \t^{ma}{}_bZ_a^A\p^{bB},\quad \bar{J}^{m}_{AB}\equiv \t^{mb}{}_a\bar{Z}^a_A\p_{bB}.
\ee

Let us now try to lift the $\CN=5$ currents (\ref{n6psc}) and (\ref{n6sccur}) to  $\CN=6$ currents. Using the above decompositions (\ref{n6fields})$-$(\ref{EFI2}), and the $USp(4)$ identity (\ref{totanti}), one can lift the $\CN=5$ action (\ref{5lagran}) to an $\CN=6$ action with manifest $SU(4)$ R-symmetry (see Ref. \cite{Hosomichi:2008jb} for details).
Specifically, the $\CN=6$ action is invariant if we apply the $SU(4)$ R-symmetry transformations to the $\CN=6$ fields:
\e\label{su4tran}\nonumber
&&\d_R\psi^{Aa}=-\frac{1}{2}\ep_{IJ}\Sigma^{IJA}_B\psi^{Ba},\quad \d_R\bar\psi_{Aa}=\frac{1}{2}\ep_{IJ}\Sigma^{IJB}_A\bar\psi_{Ba},\\
&&\d_RZ^{A}_a=-\frac{1}{2}\ep_{IJ}\Sigma^{IJA}_BZ^{B}_a,\quad
\d_R\bar Z^a_{A}=\frac{1}{2}\ep_{IJ}\Sigma^{IJB}_A\bar Z^a_{B},
\ee
where $\ep_{IJ}=-\ep_{JI}$ are set of parameters ($I,J=1,\ldots, 6$), and $\Sigma^{IJA}_B$ are the $SU(4)$ generators (see Appendix \ref{secSO6}).  As a result, the currents  $\tilde j^{I^\prime}_\mu\rightarrow\tilde j^{I^\prime}_\mu+\d_{R}\tilde j^{I^\prime}_\mu$ must be also conserved, namely,
\e\label{su4tran2}
\partial^\mu(\tilde j^{I^\prime}_\mu+\d_{R}\tilde j^{I^\prime}_\mu)=\partial^\mu(\d_{R}\tilde j^{I^\prime}_\mu)=0.
\ee
Notice that under the $SU(4)$ R-symmetry transformations (\ref{su4tran}), the currents (\ref{n6psc}) transform as
\e\label{conserved}
\d_R\tilde j^{I^\prime}_\mu=-\frac{1}{2}\ep_{KL}(\t^{KL})^{I^\prime}{}_J\tilde j^J_\mu,
\ee
where $(\t^{KL})^{I^\prime}{}_J=\d^{LI^\prime}\d^K_J-\d^{KI^\prime}\d^L_J$ are a set of  $SO(6)$ matrices, and $\tilde j^J_\mu$ are defined by replacing the $SO(5)$ gamma matrices $\Gamma^{J^\prime}_{AB}$ in $\tilde j^{J^\prime}_\mu$ (see (\ref{n6psc})) by the $SO(6)$ gamma matrices $\Gamma^{J}_{AB}=(\Gamma^{J^\prime}_{AB},i\omega_{AB})$ \footnote{The $SO(6)$ gamma matrices $\Gamma^I_{AB}$ are defined in Appendix \ref{secSO6}.}. Combining (\ref{su4tran2}) and (\ref{conserved}), one obtains
\e
\partial^\mu\tilde j^J_\mu=0.
\ee
In other words, we now have \emph{six} copies of conserved currents $\tilde j^J_\mu$; they take values in the $\textbf{6}$ of $SU(4)$. So (\ref{n6psc}) and (\ref{n6sccur}) can be promoted to the $OSp(6|4)$ currents of the $\CN=6$ theory:
\e\label{n6psc2}
(\tilde j^I_\mu)_{\CN=6}&=&-i\bar\psi_{Aa}\g_\mu(\d\psi)^{IAa}-i\psi^{Aa}\g_\mu(\d\bar\psi)^I_{Aa}
\nonumber\\&&
-\frac{i}{4}[\g_\mu,\g_\nu]\partial^\nu\big(\bar Z^a_A\bar\psi_{Ba}\Gamma^{IAB}+  Z^A_a\psi^{Ba}\Gamma^I_{AB}\big),
\\(\tilde{s}^I_\mu)_{\CN=6}&=&\g\cdot x(\tilde{j}^I_\mu)_{\CN=6},\label{n6sccur2}
\ee
where
\begin{equation}
(\d\bar\psi)^I_{Aa} = \gamma^\mu D_\mu Z^B_a\Gamma^I_{BA} +k_{mn}
  M^{mB}{}_C\t^{nb}{}_aZ^C_b\Gamma^I_{BA}+k_{mn}M^{mD}{}_A\t^{mb}{}_a
  Z^C_b \Gamma^I_{CD}
\end{equation}
is defined via the second equation of (\ref{N6susyLie}), i.e.
$\d_\ep\bar\psi_{Aa}\equiv(\d\bar\psi)^{I}_{Aa}\ep^I$;
And $(\d\psi)^{IAa}$ is the complex conjugate of $(\d\bar\psi)^I_{Aa}$.

Since there are six conserved currents $\tilde j^I_\mu$, one can also lift the $\CN=5$ super poincare transformations (\ref{n5susy}) to the ${\cal N}=6$ transformations \cite{Hosomichi:2008jb}
\begin{eqnarray}\label{N6susyLie}
\nonumber  \delta_\ep Z^A_d &=& -i\epsilon^{AB}\bar\psi_{Bd}, \\
 \nonumber
\delta_\ep \bar\psi_{Bd} &=& \gamma^\mu D_\mu Z^A_d\epsilon_{AB} +k_{mn}
  M^{mA}{}_C\t^{na}{}_dZ^C_a\epsilon_{AB}+k_{mn}M^{mD}{}_B\t^{ma}{}_d
  Z^C_a \epsilon_{CD}, \\
 \delta_\ep  A^m_\mu{}&=&
-i\epsilon_{AB}\gamma_\mu J^{mAB}+
i\epsilon^{AB}\gamma_\mu \bar{J}^m_{AB}.
\end{eqnarray}
The parameters $\epsilon_{AB}=\ep^I\Gamma^I_{AB}$ ($I=1,\ldots,6$) satisfy
\begin{equation}
\epsilon_{AB}=-\epsilon_{BA},\quad
\epsilon^*_{AB}=\epsilon^{AB}
=\frac{1}{2}\varepsilon^{ABCD}\epsilon_{CD}.
\end{equation}

Similarly, one can lift the $\CN=5$ superconformal transformations (\ref{n5csusy}) to the $\CN=6$ superconformal transformations:
\begin{eqnarray}\label{N6csusy}
\nonumber  \delta_\eta Z^A_d &=& -i\g\cdot x\eta^{AB}\psi_{Bd}, \\
 \nonumber
\delta_\eta \bar\psi_{Bd} &=& (\d\bar\psi)^I_{Bd}\g\cdot x\eta^{I}+\eta_{BA}Z_d^{A}, \\
 \delta_\eta  A^m_\mu{}&=&
-i(\g\cdot x\eta_{AB})\gamma_\mu J^{mAB}+
i(\g\cdot x\eta^{AB})\gamma_\mu \bar{J}^m_{AB}.
\end{eqnarray}
The set of parameters $\eta_{AB}=\eta^I\Gamma^I_{AB}$ ($I=1,\ldots,6$) obey the anti-symmetry and reality conditions:
\begin{equation}
\eta_{AB}=-\eta_{BA},\quad
\eta^*_{AB}=\eta^{AB}
=\frac{1}{2}\varepsilon^{ABCD}\eta_{CD}.
\end{equation}

Finally, we would like to comment on the fundamental identity (\ref{EFI2}).
It can be understood as the $Q\bar QQ$ or  $\bar QQ\bar Q$ Jacobi identity of the
superalgebra \cite{Chen:pku1}:
\e\label{SLie6} &&[M^m, M^n]=C^{mn}{}_sM^s,\nonumber\\
&&[M^m, Q^a]=-\t^{ma}{}_bQ^b,\quad [M^m, \bar
Q_a]=\t^{mb}{}_a\bar Q_b,\nonumber\\
&&\{Q^a, \bar Q_{b}\}=\t^{ma}{}_bk_{mn}M^n,\quad \{\bar Q_{a}, \bar
Q_b\}=\{Q^{a}, Q^b\}=0.\ee
So we must select the bosonic subalgebra of (\ref{SLie6}) as the Lie algebra of the
gauge group. One can specify (\ref{SLie6}) as one of the superalgebras  \cite{Hosomichi:2008jb}
\e
OSp(2|2N),\quad U(M|N),\quad SU(M|N), \quad PSU(M|N).
\ee
For instance, if we select the bosonic subalgebra of $U(N|N)$ as the Lie algebra of the gauge symmetry, then the theory will become the well known $\CN=6$ ABJM theory \cite{ABJM}. In particular, if we specify (\ref{SLie6}) as $PSU(2|2)$, whose bosonic part is $SU(2)\times SU(2)$, the supersymmetry will be enhanced to $\CN=8$ \cite{HosomichiJD,ABJM}, and the theory will become the well known $\CN=8$ BLG theory \cite{Bagger,Gustavsson}. So the $OSp(8|4)$ superconformal currents, which were first constructed by studying the $\CN=8$ BLG theory directly \cite{Schwarz0}, should be also derived as specific examples of the currents (\ref{n6psc}) and (\ref{n6sccur}) associated with the general $\CN=6$ theory.

\subsection{Currents in the ABJM Theories}
The $\CN=6$ super poincare and superconformal currents of the ABJM theory were first derived in Ref. \cite{Schwarz}, by studying the ABJM theory derictly. In this section, we will show that they can be  re-derived as specific examples of the $OSp(6|4)$ currents (\ref{n6psc}) and (\ref{n6sccur}) associated with the general $\CN=6$ theory. To see this, let us first specify the superalgebra (\ref{SLie6}) as $U(M|N)$. Or in other words, we choose the bosonic subalgebra of $U(M|N)$ as the Lie algebra of the gauge symmetry. So the gauge group is $U(M)\times U(N)$. If $M=N$, it becomes the well-known ABJM theory \cite{ABJM}. The ABJM theory has been conjectured to be the gauge description of M2-branes probing a $\textbf{C}^4/\textbf{Z}_k$ singularity.

We denote the scalar field its complex conjugate as $Z_{n\hat{n}}^A$ and $\bar Z^{\hat{n}n}_A$, respectively. Here $n=1,...,M$ is a fundamental index of
$U(M)$, $\hat n= 1,...,N$ an anti-fundamental index of $U(N)$. The fermionic field and its complex conjugate are denoted as $\psi^{A\hat{n}n}$ and $\bar\psi_{An\hat{n}}$, respectively.
The $U(M)$  and
$U(N)$ parts of the gauge potential are defined as $\hat{A}_\mu{}^{\hat k}{}_{\hat n}$ and $A_\mu{}^{k}{}_{n}$, respectively.
With
these notations, the hermitian inner product of two matter fields
can be written as a trace:
\begin{equation}
\bar{X}^{\hat nn}Y_{n\hat
n}\equiv {\rm Tr}(\bar{X}Y).
\end{equation}

In order to compare our results with that of Ref. \cite{Schwarz}, from now on, we assume that both $n$ and $\hat{n}$ run from $1$ to $N$.

In matrix notation, the ${\cal N}=6$ super poincare transformations (\ref{N6susyLie}) for the $U(N)\times U(N)$ theory are given by \footnote{We refer the reader to Ref. \cite{Hosomichi:2008jb} for the computational details.}:
\begin{eqnarray}\label{UnitarySusyTrasf}
\nonumber  \delta_\ep Z^A &=& -i\epsilon^{AB}\bar\psi_{B}, \\
 \nonumber
  \delta_\ep\bar\psi_{B} &=& \gamma^\mu D_\mu Z^A\epsilon_{AB} +
k(Z^C\bar{Z}_C Z^A-Z^A\bar{Z}_C
Z^C)\epsilon_{AB}+2kZ^C\bar{Z}_BZ^D\epsilon_{CD},
\nonumber\\
\delta_\ep\hat{A}_\mu&=&-ik\epsilon_{AB}\gamma_\mu\psi^B
Z^A+ik\epsilon^{AB}\gamma_\mu\bar{Z}_A\bar\psi_B,\\
\delta_\ep A_\mu&=&ik\epsilon_{AB}\gamma_\mu Z^A\psi^B
-ik\epsilon^{AB}\gamma_\mu\bar\psi_B\bar{Z}_A,\nonumber
\end{eqnarray}
where $k$ is a constant. Writting the second equation as $\delta_\ep\bar\psi_{B}=(\delta\bar\psi)^I_{B}\ep^I$, with
\e
(\delta\bar\psi)^I_{B} = \gamma^\mu D_\mu Z^A\Gamma^I_{AB} +
k(Z^C\bar{Z}_C Z^A-Z^A\bar{Z}_C
Z^C)\Gamma^I_{AB}+2kZ^C\bar{Z}_BZ^D\Gamma^I_{CD},
\ee
the superconformal transformations (\ref{N6csusy}) now read:
\begin{eqnarray}\label{N6csusy2}
\nonumber  \delta_\eta Z^A &=& -i\g\cdot x\eta^{AB}\psi_{B}, \\
 \nonumber
\delta_\eta \bar\psi_{B} &=& (\d\bar\psi)^I_{Bd}\g\cdot x\eta^{I}+\eta_{BA}Z^{A}, \\
\delta_\eta\hat{A}_\mu&=&-ik(\g\cdot x\eta_{AB})\gamma_\mu\psi^B
Z^A+ik(\g\cdot x\eta^{AB})\gamma_\mu\bar{Z}_A\bar\psi_B,\\
\delta_\eta A_\mu&=&ik(\g\cdot x\eta_{AB})\gamma_\mu Z^A\psi^B
-ik(\g\cdot x\eta^{AB})\gamma_\mu\bar\psi_B\bar{Z}_A.\nonumber
\end{eqnarray}

And the currents (\ref{n6psc}) and (\ref{n6sccur}) now become
\e\label{n6psc2}
(\tilde j^I_\mu)_{ABJM}&=&-i{\rm Tr}[\bar\psi_{A}\g_\mu(\d\psi)^{IA}]-i{\rm Tr}[\psi^{A}\g_\mu(\d\bar\psi)^I_{A}]
\nonumber\\&&
-\frac{i}{4}[\g_\mu,\g_\nu]\partial^\nu\big[{\rm Tr}(\bar Z_A\bar\psi_{B})\Gamma^{IAB}+ {\rm Tr}(Z^A\psi^{B})\Gamma^I_{AB}\big],
\\ (\tilde{s}^I_\mu)_{ABJM}&=&\g\cdot x(\tilde{j}^I_\mu)_{ABJM},\label{n6sccur3}
\ee
Define
\e
&&(j^I_\mu)_{ABJM}=-i{\rm Tr}[\bar\psi_{A}\g_\mu(\d\psi)^{IA}]-i{\rm Tr}[\psi^{A}\g_\mu(\d\bar\psi)^I_{A}],\\
&&B^I_{\mu\nu}\equiv -\frac{i}{4}[\g_\mu,\g_\nu]\big[{\rm Tr}(\bar Z_A\bar\psi_{B})\Gamma^{IAB}+ {\rm Tr}(Z^A\psi^{B})\Gamma^I_{AB}\big]=-B^I_{\nu\mu},\label{Bmunu}
\ee
then (\ref{n6psc2}) and (\ref{n6sccur3}) can be written as
\e\label{n6psc3}
(\tilde j^I_\mu)_{ABJM}&=&(j^I_\mu)_{ABJM}+\partial^\nu(B^I_{\mu\nu})\\
(\tilde{s}^I_\mu)_{ABJM}&=&\g\cdot x(j^I_\mu)_{ABJM}-i\big[{\rm Tr}(\bar Z_A\g_\mu\bar\psi_{B})\Gamma^{IAB}+ {\rm Tr}(Z^A\g_\mu\psi^{B})\Gamma^I_{AB}\big]\nonumber\\
&&+\partial^{\nu}(\g\cdot xB^I_{\mu\nu}).\label{n6sccur4}
\ee
The above $OSp(6|4)$ currents for the ABJM theory are in agreement\footnote{Notice that our reality condition is $\epsilon^*_{AB}=\epsilon^{AB}
=\frac{1}{2}\varepsilon^{ABCD}\epsilon_{CD}$, while in \cite{Schwarz}, it is defined by $\epsilon^\dag_{AB}=\epsilon^{AB}
=\frac{1}{2}\varepsilon^{ABCD}\epsilon_{CD}$.} with the ones constructed in Ref. \cite{Schwarz}, since they are only up \emph{conserved total derivative terms}.
Indeed, both the last term of (\ref{n6psc3}) and the last term of (\ref{n6sccur3}) are total derivative terms, and satisfy $\partial^\mu\partial^\nu(B^I_{\mu\nu})=\partial^\mu\partial^{\nu}(\g\cdot xB^I_{\mu\nu})=0$ on account of the fact that $B^I_{\mu\nu}=-B^I_{\nu\mu}$.

Similarly, one can also derive the $OSp(6|4)$ currents associated with the ${\cal
N}=6, Sp(2N)\times U(1)$ theory \cite{Hosomichi:2008jb,ChenWu1}.

\section{Closure of the $OSp(5|4)$ Superconformal algebra}\label{SecClosure}
The closure of the $\CN=5$ super poincare algebra was checked in our previous work \cite{Chen2}. In this section we will check the closure of the full $OSp(5|4)$ superconformal algebra.

Let us begin by considering the scalar fields. In Ref. \cite{Chen2}, it was shown that\footnote{In Ref. \cite{Chen2}, we worked in the symplectic 3-algebra framework. Here we present a Lie algebra version of the closure of the super poincare algebra, by converting the 3-algebra description into the conventional Lie algebra, using the method described in Ref. \cite{ChenWu3}. The key point is that the 3-algebra structure constants $f_{abc}{}^d$ can be constructed in terms of the tensor product on the superalgebra (\ref{SLie}): $f_{abc}{}^d=k_{mn}\t^m_{ab}\t^{nd}{}_c$ \cite{ChenWu3}. (One may also use certain curvature tensors to construct the 3-algebras \cite{chen5}.)}
\begin{eqnarray}\label{commZ1}
[\delta_{\ep_1}, \delta_{\ep_2}]Z^{a}_{A}=v_1^{\mu}D_{\mu}Z^{a}_{A}+\tilde{\Lambda}_1^a{}_bZ^b_A,
\end{eqnarray}
where $\delta_{\ep_1}$ and $\delta_{\ep_2}$ are two super poincare transformations, and
\begin{eqnarray}
v_1^{\mu} &\equiv& -\frac{i}{2}\epsilon_{2}^{BD}\gamma^{\mu}
\epsilon_{1BD}=-2i\ep_2^I\g^\mu\ep_1^I,\\
\tilde{\Lambda}_1^a{}_b&\equiv&\Lambda^{cd}_1k_{mn}\t^m_{cd}\t^{na}{}_b, \\
\Lambda^{cd}_1&\equiv& -\frac{i}{2}Z^{c}_DZ^{d}_{C}
(\epsilon_{1}^{CE}\epsilon_{2E}{}^D-\epsilon_{2}^{CE}
\epsilon_{1E}{}^D)=-2iZ^{c}_DZ^{d}_{C}\Gamma^{IJCD}\ep_2^I\ep_1^J\label{lambda}
\end{eqnarray}
where $I=1,\ldots,5$ is a fundamental index of $SO(5)$, and
\e
\Gamma^{IJCD}=\frac{1}{4}(\Gamma^{ICA}\Gamma^J_A{}^D-\Gamma^{JCA}\Gamma^I_A{}^D)
\ee
are a set of $USp(4)$ generators, with the $SO(5)$ matrices $\Gamma^J_A{}^D$ defined in Appendix \ref{secSO5}. Notice that the first term of
Eq. (\ref{commZ1}) is the gauge covariant translation, while the second term is a gauge transformation.

Having (\ref{commZ1}), it is much easier to evaluate the commutator
\begin{eqnarray}\
[\delta_{\eta_1}, \delta_{\ep_2}]Z^{a}_{A},
\end{eqnarray}
where $\delta_{\eta_1}$ is an $\CN=5$ superconformal transformation defined by (\ref{n5csusy}). If fact, following the idea of Ref. \cite{Schwarz}, we have shown that the superconformal transformations (\ref{n5csusy}) can be obtained from the poincare transformations (\ref{n5susy}) by replacing the parameters $\ep$ with $ \g\cdot x\eta$ and adding \e\label{atranpsi2}
\d^\prime_{\eta_1}\psi^a_A=-\eta_{1A}{}^BZ^a_B
\ee
into the variation of the fermion fields (see Section \ref{SecN51}). Using this
idea and Eq. (\ref{commZ1}), we have immediately
\begin{eqnarray}\label{commZ2}
[\delta_{\eta_1}, \delta_{\ep_2}]Z^{a}_{A}=v_2^{\mu}D_{\mu}Z^{a}_{A}+\tilde{\Lambda}_2^a{}_bZ^b_A
-i\ep_{2A}{}^B\eta_{1B}{}^CZ^a_C.
\end{eqnarray}
Here $v_2$ is obtained from $v_1$ by the replacing $\ep_1$ in $v_1$ with $\g\cdot x\eta_1$, and $\tilde{\Lambda}_2^a{}_b$ is obtained from $\tilde{\Lambda}_1^a{}_b$ by the replacing $\ep_1$ in $\tilde{\Lambda}_1^a{}_b$ with $\g\cdot x\eta_1$, i.e.
\e
v_2^{\mu} &\equiv& -2i\ep_2^I\g^\mu(\g\cdot x\eta_1^I)\label{v2}\\
\tilde{\Lambda}_2^a{}_b&\equiv&\Lambda^{cd}_2k_{mn}\t^m_{cd}\t^{na}{}_b=
\Lambda^m_2k_{mn}\t^{na}{}_b, \label{lambda2}\\
\Lambda^{cd}_2&\equiv&-2iZ^{c}_DZ^{d}_{C}\Gamma^{IJCD}\ep_2^I(\g\cdot x\eta_1^J).
\ee
The last term of (\ref{commZ2}) is a direct consequence of (\ref{atranpsi2}).
Notice that Eq. (\ref{commZ2}) is a gauge covariant equation. On the other hand, after some algebra, we find that (\ref{atranpsi2}) can be written as
\e
[\delta_{\eta_1}, \delta_{\ep_2}]Z^{a}_{A}&=&-2i(\ep^I_2\eta^I_1)(x^\mu\partial_\mu+\frac{1}{2})Z^a_A-
2(\ep^I_2\g^{\mu\nu}\eta_1^I)(x_\mu\partial_\nu-x_\nu\partial_\mu)Z^a_A\nonumber\\
&&-2i(\ep^I_2\eta^J_1)\Gamma^{IJ}_A{}^BZ^a_B+(\tilde\Lambda_2^a{}_b+v^\mu_2A^m_\mu\t_m{}^a{}_b)Z^b_A,
\label{commZ4}
\ee
where
\e\label{gammamunu2}
\g^{\mu\nu}=-\frac{i}{4}[\g^\mu, \g^\nu].
\ee
The first three terms are scale, Lorentz, and $USp(4)$ R-symmetry transformations, respectively; the last term is a gauge transformation by the parameter $(\tilde\Lambda_2^a{}_b+v^\mu_2A^m_\mu\t_m{}^a{}_b)$. In particular, the first term indicates that the dimension of $Z^a_A$ is $\frac{1}{2}$.

Similarly, applying the replacement $\ep_2\rightarrow \g\cdot x\eta_2$ to the RHS of (\ref{commZ2}), and taking account of (\ref{atranpsi2}), we obtain
\begin{equation}\label{commZ4}
[\delta_{\eta_1}, \delta_{\eta_2}]Z^{a}_{A}=v_3^{\mu}D_{\mu}Z^{a}_{A}+\tilde{\Lambda}_3^a{}_bZ^b_A
-i(\g\cdot x\eta_{2A}{}^B)\eta_{1B}{}^CZ^a_C+i(\g\cdot x\eta_{1A}{}^B)\eta_{2B}{}^CZ^a_C,
\end{equation}
where
\e
v_3^{\mu} &\equiv& -2i[(\g\cdot x\eta_2^I)\g^\mu(\g\cdot x\eta_1^I)]\label{v3}\\
\tilde{\Lambda}_3^a{}_b&\equiv&\Lambda^{cd}_3k_{mn}\t^m_{cd}\t^{na}{}_b=\Lambda^m_3k_{mn}\t^{na}{}_b,\label{lambda3} \\
\Lambda^{cd}_3&\equiv&-2iZ^{c}_DZ^{d}_{C}\Gamma^{IJCD}[(\g\cdot x\eta_2^I)(\g\cdot x\eta_1^J)].
\ee
Notice that Eq. (\ref{commZ4}) is manifestly gauge covariant. Also, after some work, we obtain
\e\label{commZ5}
[\delta_{\eta_1}, \delta_{\eta_2}]Z^{a}_{A}&=&2(\eta^I_1\g^\nu\eta^I_2)[(-2ix_\nu x^\mu\partial_\mu+ix^2\partial_\nu)Z^a_A-ix_\nu Z^a_A]\nonumber\\&&+(\tilde\Lambda_3^a{}_b+v^\mu_3\tilde A_\mu^a{}_b)Z^b_A.
\ee
We see that the first line is the standard special superconformal transformation, while the second line is just a gauge transformation.

Let us now check the gauge fields. In Ref. \cite{Chen2}, it was shown that
\begin{eqnarray}\label{A1}
[\delta_{\ep_1}, \delta_{\ep_2}]A^m_\mu&=& v_1^\nu F^m_{\nu\mu}-D_\mu\Lambda^m_1 \nonumber\\
&&+v_1^\nu[F^m_{\mu\nu}-\varepsilon_{\mu\nu\lambda}(Z^a_AD^\lambda Z^{Ab}-\frac{i}{2}\bar{\psi}^{Ba}\gamma^\lambda\psi^b_B)\t^m_{ab}],
\end{eqnarray}
where $\Lambda^m_1=\Lambda^{ab}_1\t^m_{ab}$, and $\Lambda^{ab}_1$ is defined by (\ref{lambda}).
The second term of the first line is a gauge transformation. And the second line is the equations of motion for the gauge fields. In order to use the $\ep_1\rightarrow \g\cdot x\eta_1$ trick, we must rewrite (\ref{A1}) as
\begin{eqnarray}\label{A2}
[\delta_{\ep_1}, \delta_{\ep_2}]A^m_\mu&=& v_1^\nu F^m_{\nu\mu}-(\ep_2^I\ep_1^J)D_\mu(-2iZ^{a}_AZ^{b}_{B}\Gamma^{IJAB}\t^m_{ab}) +{\rm EOM},
\end{eqnarray}
since $\ep_2^I$ and $\ep_1^J$ are \emph{constant parameters}. (``EOM" stands for ``equations of motion".) Applying the replacement $\ep_1\rightarrow \g\cdot x\eta_1$ to the RHS of (\ref{A2}), and tacking account of (\ref{atranpsi2}), one obtains
\begin{eqnarray}\label{A3}
[\delta_{\eta_1}, \delta_{\ep_2}]A^m_\mu&=& v_2^\nu F^m_{\nu\mu}-(\ep_2^I\g\cdot x\eta_1^J)D_\mu(-2iZ^{a}_AZ^{b}_{B}\Gamma^{IJAB}\t^m_{ab})\nonumber\\
&&+i[\ep^{AB}_2\g_\mu(-\eta_{1B}{}^CZ^b_C)]Z^a_A\t^m_{ab}  +{\rm EOM}\nonumber\\
&=& v_2^\nu F^m_{\nu\mu}-D_\mu\Lambda^m_2 +{\rm EOM},
\end{eqnarray}
where $\Lambda^m_2$ is defined by (\ref{lambda2}). It can be recast into the form
\begin{eqnarray}\label{A4}
[\delta_{\eta_1}, \delta_{\ep_2}]A^m_\mu
&=&-2i(\ep^I_2\eta^I_1)(x^\mu\partial_\mu+1)A^m_\mu\nonumber\\&&-
2i(\ep^I_2\g^{\rho\sigma}\eta_1^I)[-i(x_\rho\partial_\sigma-x_\sigma\partial_\rho)A^m_\mu
+(S_{\rho\sigma})_\mu{}^\nu A^m_\nu]
\nonumber\\&&-D_\mu(\Lambda^m_2+v^\nu_2 A^m_\nu) +{\rm EOM},
\end{eqnarray}
where
\e\label{lorentz}
(S_{\rho\sigma})_\mu{}^\nu=i(\eta_{\sigma\mu}\d^\nu_\rho-
\eta_{\rho\mu}\d^\nu_\sigma).
\ee
The first and second lines are scale and Lorentz transformations respectively, while
the first term of the third line is a gauge transformation by the parameter $(\Lambda^m_2+v^\nu_2 A^m_\nu)$, as expected. The first line indicates that the dimension of $A^m_\mu$ is $1$.

Similarly, \emph{after} writing $-D_\mu\Lambda^m_2$ in (\ref{A3}) as
\e
-D_\mu\Lambda^m_2=-(\ep_2^I\g_\nu\eta_1^J)D_\mu(-2ix^\nu Z^{c}_DZ^{d}_{C}\Gamma^{IJCD}\t^m_{ab}),
\ee
one can calculate $[\delta_{\eta_1}, \delta_{\eta_2}]A^m_\mu$ by applying the replacement $\ep_2\rightarrow \g\cdot x\eta_2$ to the RHS of (\ref{A3}), while taking account of (\ref{atranpsi2}). After some algebraic steps, we obtain the desired results
\begin{eqnarray}\label{A5}
&&[\delta_{\eta_1}, \delta_{\eta_2}]A^m_\mu\nonumber\\
&=& v_3^\nu F^m_{\nu\mu}-D_\mu\Lambda^m_3\\
&=&2(\eta^I_1\g^\nu\eta^I_2)[(-2ix_\nu x^\rho\partial_\rho+ix^2\partial_\nu)A^m_\mu-2ix_\nu A^m_\mu-2x^\rho(S_{\rho\nu})_\mu{}^\sigma A^m_\sigma)]\label{A6}\\&&
-D_\mu(\Lambda^m_3+v^\nu_3A^m_\mu) +{\rm EOM},\nonumber
\end{eqnarray}
where $v^\nu_3$, $\Lambda^m_3$, and $(S_{\rho\nu})_\mu{}^\sigma$ are defined by (\ref{v3}), (\ref{lambda3}), and (\ref{lorentz}), respectively. It can be seen that
the first line of (\ref{A6}) is the standard special conformal transformation.

Finally, we examine the fermion fields. In Ref. \cite{Chen2}, it has been checked that
\begin{eqnarray}\label{SusyOnPsi}
\nonumber [\delta_{\ep_1},\delta_{\ep_2}]\psi^a_{A} &=& v_1^\mu D_\mu
\psi^a_{A} + \tilde{\Lambda}_1^a{}_{b}
\psi^b_{A}\\
\nonumber &&+\frac{i}{2}(\epsilon_1^{BC}\epsilon_{2BA}
-\epsilon_2^{BC}\epsilon_{1BA})E^a_{C}\\
 &&
-\frac{1}{2}v_1^\nu\gamma_\nu E^a_{A}.
\end{eqnarray}
The last two lines are the equations of motion for fermionic fields, i.e.  $E^a_{A}=0$, where
\begin{equation}
E^a_{A} = \gamma^\mu D_\mu\psi^a_{A}
-k_{mn}\t^m_{cd}\t^{na}{}_bZ^b_BZ^{Bc}\psi^d_A+2k_{mn}\t^m_{cd}\t^{na}{}_bZ^b_BZ^c_A\psi^{Bd}.
\end{equation}

One can examine $[\delta_{\eta_1},\delta_{\ep_2}]\psi^a_{A}$ and $[\delta_{\eta_1},\delta_{\eta_2}]\psi^a_{A}$, using the same strategy applied to   the scalar and gauge fields. First, we have
\begin{eqnarray}\label{SusyOnPsi2}
\nonumber [\delta_{\eta_1},\delta_{\ep_2}]\psi^a_{A} &=& v_2^\mu D_\mu
\psi^a_{A} + \tilde{\Lambda}_2^a{}_{b}
\psi^b_{A}\nonumber\\&&
-i(\psi^a_C\g^\mu\eta_{1B}{}^C)(\g_\mu\ep_{2A}{}^B)+i\eta_{1A}{}^B(\ep_{2B}{}^C\psi^a_C)
\\&&+ {\rm EOM}\nonumber
\end{eqnarray}
Using the Fierz transformations (\ref{Fierz}), one can recast the above equation as
\begin{eqnarray}
[\delta_{\eta_1},\delta_{\ep_2}]\psi^a_{A}&=&-2i(\ep^I_2\eta^I_1)(x^\mu\partial_\mu+1)\psi^a_A
\nonumber\\&&-
2(\ep^I_2\g^{\mu\nu}\eta_1^I)(x_\mu\partial_\nu-x_\nu\partial_\mu+i\g_{\mu\nu})\psi^a_A\nonumber\\
&&-2i(\ep^I_2\eta^J_1)\Gamma^{IJ}_A{}^B\psi^a_B\nonumber\\&&
+(\tilde\Lambda_2^a{}_b+v^\mu_2A^m_\mu\t_m{}^a{}_b)\psi^b_A+{\rm EOM}.
\end{eqnarray}
The first three lines are scale, Lorentz, and $USp(4)$ R-symmetry transformations, while the last line is a gauge transformation and the equations of motion. The first line indicates that the dimension of the fermion field is $1$.

Finally, we have
\e\label{psi4}
&&[\delta_{\eta_1}, \delta_{\eta_2}]\psi^{a}_{A}\nonumber\\
&=&2(\eta^I_1\g^\nu\eta^I_2)[(-2ix_\nu x^\mu\partial_\mu+ix^2\partial_\nu)Z^a_A+(-2ix_\nu -2x^\mu\g_{\mu\nu})\psi^a_A]\nonumber\\&&+(\tilde\Lambda_3^a{}_b+v^\mu_3\tilde A_\mu^a{}_b)\psi^b_A\nonumber\\&&+{\rm EOM}.
\ee
It can be seen that the first line is the special conformal transformation, while the second line is a gauge transformation. Using the fact that $v^\mu_3$ can be written as
\e
v^\mu_3=-4i(\eta^I_1\g^\nu\eta^I_2)x_\nu x^\mu+2i(\eta^I_1\g^\mu\eta^I_2)x^2,
\ee
Eq. (\ref{psi4}) can be convert into the manifestly covariant form:
\e\label{psi5}
&&[\delta_{\eta_1}, \delta_{\eta_2}]\psi^{a}_{A}\nonumber\\
&=&2(\eta^I_1\g^\nu\eta^I_2)[(-2ix_\nu x^\mu D_\mu+ix^2D_\nu)Z^a_A+(-2ix_\nu -2x^\mu\g_{\mu\nu})\psi^a_A]\nonumber\\&&+\tilde\Lambda_3^a{}_b\psi^b_A\nonumber\\&&+{\rm EOM}.
\ee

\section{Conclusions}\label{SecDis}
In this paper, we have verified that the general $\CN=5$ theory has a complete $OSp(5|4)$ superconformal symmetry, and constructed the corresponding conserved supercurrents; we have also derived the super currents in the $\CN=5$, $USp(2N)\times SO(M)$ as special examples. The $OSp(5|4)$ superconformal algebra was shown to be closed on shell. These $OSp(5|4)$ superconformal currents may be useful in constructing the M2-brane superconformal algebras, and in studying the BPS brane configurations.

We have demonstrated that the $OSp(6|4)$ superconformal currents associated with the general $\CN=6$ theory can be obtained as special cases of the $OSp(5|4)$ superconformal currents by promoting the $USp(4)$ R-symmetry to $SU(4)$. Specifying the gauge group as $U(N)\times U(N)$, we have been able to rederive the $OSp(6|4)$ superconformal currents associated with the ABJM theory which were first constructed in Ref .\cite{Schwarz}.

We have also argued that the $OSp(8|4)$ superconformal currents associated with the $\CN=8$ BLG theory \cite{Schwarz0} can be re-derived as the special cases of the  general $OSp(6|4)$ currents by enhancing the $SU(4)$ R-symmetry to $SO(8)$. However, it would be nice to prove this relation explicitly.

In this paper, we have worked in an ordinary Lie algebra framework. However, it would be nice to construct all these supercurrents in a 3-algebra framework.

\section{Acknowledgement}
we thank the referee for useful comments. We thank for the Korea Institute for Advanced Study for warm hospitality during the beginning stage of
this work. This work is supported by the China Postdoctoral Science Foundation through Grant No. 2012M510244.

\appendix

\section{Conventions and Useful Identities}\label{Identities}

The conventions and identities of this appendix are mainly adopted from Ref. \cite{ChenWu3}.
\subsection{Spinor Algebra}
In $1+2$ dimensions, the gamma matrices are defined as
\begin{equation}
(\gamma_{\mu})_{\alpha}{}^\gamma(\gamma_{\nu})_{\gamma}{}^\beta+
(\gamma_{\nu})_{\alpha}{}^\gamma(\gamma_{\mu})_{\gamma}{}^\beta=
2\eta_{\mu\nu}\delta_{\alpha}{}^\beta.
\end{equation} For the metric we
use the $(-,+,+)$ convention. The gamma matrices in the Majorana
representation can be defined in terms of Pauli matrices:
$(\gamma_{\mu})_{\alpha}{}^\beta=(i\sigma_2, \sigma_1, \sigma_3)$,
satisfying the important identity
\begin{equation}
(\gamma_{\mu})_{\alpha}{}^\gamma(\gamma_{\nu})_{\gamma}{}^\beta
=\eta_{\mu\nu}\delta_{\alpha}{}^\beta+\varepsilon_{\mu\nu\lambda}(\gamma^{\lambda})_{\alpha}{}^\beta.
\end{equation}
We also define
$\varepsilon^{\mu\nu\lambda}=-\varepsilon_{\mu\nu\lambda}$. So
$\varepsilon_{\mu\nu\lambda}\varepsilon^{\rho\nu\lambda} =
-2\delta_\mu{}^\rho$. We raise and lower spinor indices with an
antisymmetric matrix
$\epsilon_{\alpha\beta}=-\epsilon^{\alpha\beta}$, with
$\epsilon_{12}=-1$. For example,
$\psi^\alpha=\epsilon^{\alpha\beta}\psi_\beta$ and
$\gamma^\mu_{\alpha\beta}=\epsilon_{\beta\gamma}(\gamma^\mu)_\alpha{}^\gamma
$, where $\psi_\beta$ is a Majorana spinor. Notice that
$\gamma^\mu_{\alpha\beta}=(\mathbbm{l}, -\sigma^3, \sigma^1)$ are
symmetric in $\alpha\beta$. A vector can be represented by a
symmetric bispinor and vice versa:
\begin{equation}
A_{\alpha\beta}=A_\mu\gamma^\mu_{\alpha\beta},\quad\quad A_\mu=-\frac{1}{2}\gamma^{\alpha\beta}_\mu A_{\alpha\beta}.
\end{equation}
We use the following spinor summation convention:
\begin{equation}
\psi\chi=\psi^\alpha\chi_\alpha,\quad\quad
\psi\gamma_\mu\chi=\psi^\alpha(\gamma_{\mu})_{\alpha}{}^\beta\chi_\beta,
\end{equation}
where $\psi$ and $\chi$ are anti-commuting Majorana spinors. In
$1+2$ dimensions the Fierz transformations are
\begin{eqnarray}\label{Fierz}
(\lambda\chi)\psi &=& -\frac{1}{2}(\lambda\psi)\chi -\frac{1}{2}
(\lambda\gamma_\nu\psi)\gamma^\nu\chi.
\end{eqnarray}
Finally, we define
\e\label{gammamunu1}
\g^{\mu\nu}=-\frac{i}{4}[\g^\mu, \g^\nu].
\ee

\subsection{$SO(4)$, $SO(5)$, and $SO(6)$ Gamma Matrices}\label{SecSOM}
\subsubsection{$SO(4)$ Gamma Matrices}
We define the 4 sigma matrices as
\begin{equation}\label{pulim}
\sigma^a{}_A{}^{\dot{B}}=(\sigma^1,\sigma^2,\sigma^3,i\mathbbm{l}),
\end{equation}
by which one can establish a connection between the $SU(2)\times
SU(2)$ and $SO(4)$ group. These sigma matrices satisfy the following
Clifford algebra:
\begin{eqnarray}
\sigma^a{}_{A}{}^{\dot{C}}\sigma^{b\dag}{}_{\dot{C}}{}^B+
\sigma^b{}_{A}{}^{\dot{C}}\sigma^{a\dag}{}_{\dot{C}}{}^B=2\delta^{ab}\delta_A{}^B,\\
\sigma^{a\dag}{}_{\dot{A}}{}^{C}\sigma^{b}{}_{C}{}^{\dot{B}}+
\sigma^{b\dag}{}_{\dot{A}}{}^{C}\sigma^{a}{}_{C}{}^{\dot{B}}=2\delta^{ab}\delta_{\dot{A}}{}^{\dot{B}}.
\end{eqnarray}
We use antisymmetric matrices
\begin{eqnarray}
\epsilon_{AB}=-\epsilon^{AB}=\begin{pmatrix} 0&-1 \\ 1&0
\end{pmatrix}\;
\;\;{\rm and}\;\;\;
\epsilon_{\dot{A}\dot{B}}=-\epsilon^{\dot{A}\dot{B}}=\begin{pmatrix}
0&1 \\-1& 0
\end{pmatrix}
\end{eqnarray}
to raise or lower un-dotted and dotted indices, respectively. For
example,
$\sigma^{a\dag\dot{A}B}=\epsilon^{\dot{A}\dot{B}}\sigma^{a\dag}{}_{\dot{B}}{}^{B}$
and $\sigma^{aB\dot{A}}=\epsilon^{BC}\sigma^{a}{}_{C}{}^{\dot{A}}$.
The sigma matrix $\sigma^a$ satisfies a reality condition
\begin{equation}\label{RC4}
\sigma^{a\dag}{}_{\dot{A}}{}^{B}=-\epsilon^{BC}\epsilon_{\dot{A}\dot{B}}\sigma^a{}_{C}{}^{\dot{B}},\quad
{\rm or} \quad\sigma^{a\dag\dot{A}B}=-\sigma^{aB\dot{A}}.
\end{equation}

\subsubsection{$SO(5)$ Gamma Matrices}\label{secSO5}

We define the $SO(5)$ gamma matrices as \footnote{To avoid introducing too many indices, we still use the capital letters $A, B, \ldots$ to
label the $USp(4)$ indices. We hope this will not cause any confusion.}
\begin{eqnarray}\label{5Gamma}
\Gamma^a_A{}^B=\begin{pmatrix}0&\sigma^a\\\sigma^{a\dag}&0
\end{pmatrix},\quad\quad
\Gamma^5_A{}^B=(\Gamma^1\Gamma^2\Gamma^3\Gamma^4)_A{}^B.
\end{eqnarray}
Notice that
$\Gamma^I_A{}^B$ ($I=1,\ldots, 5$) are Hermitian, satisfying the
Clifford algebra
\begin{equation}\label{so5g}
\Gamma^{I}_A{}^C\Gamma^{J}_C{}^B+\Gamma^{J}_A{}^C\Gamma^{I}_C{}^B
=2\delta^{IJ}\delta_A{}^B.
\end{equation}
We use an antisymmetric matrix $\omega_{AB}=-\omega^{AB}$ to lower
and raise indices; for instance
\begin{equation}\label{5raise}
\Gamma^{IAB}=\omega^{AC}\Gamma^I_C{}^B.
\end{equation}
It can be chosen as the charge conjugate matrix:
\begin{equation}\label{antiform}
\omega^{AB}=\begin{pmatrix} \epsilon^{AB} & 0 \\
0 & \epsilon^{\dot{A}\dot{B}}
\end{pmatrix}.
\end{equation}
(Recall that $A$ and $\dot{B}$ of the RHS run from 1 to 2.)

By the definition (\ref{5Gamma}) and the convention (\ref{5raise}),
the gamma matrix $\Gamma^I$ is antisymmetric and traceless, and
satisfies a reality condition
\begin{eqnarray}
\Gamma^{IAB}=-\Gamma^{IBA} \quad,\quad \Gamma^{I}_A{}^A=0\quad {\rm
and}\quad
\Gamma^{I*}_{AB}=\Gamma^{IAB}=\omega^{AC}\omega^{BD}\Gamma^I_{CD}.
\end{eqnarray}

There is a useful $USp(4)$ identity
\begin{eqnarray}\label{totanti}
\varepsilon^{ABCD}&=&-\omega^{AB}\omega^{CD}
+\omega^{AC}\omega^{BD}-\omega^{AD}\omega^{BC}.
\end{eqnarray}

\subsubsection{$SO(6)$ Gamma Matrices}\label{secSO6}
Define the first five gamma matrices as the $SO(5)$ gamma matrices (\ref{so5g}), and define $\Gamma^6_{AB}=i\omega_{AB}$ and $\Gamma^{6AB}=i\omega^{AB}$ (see (\ref{antiform}) for the definition of $\omega^{AB}$), we then have the Clifford algebra\footnote{In order to avoid introducing too many indices, here we still use $I$ to label the $SO(6)$ indices, while in Appendix \ref{secSO5}, we use it to label the $SO(5)$ indices. We hope this will not cause any confusion.}
\e
\Gamma^I_{AB}\Gamma^{JBC}+\Gamma^J_{AB}\Gamma^{IBC}=-2\d^{IJ}\d^C_A,
\ee
where $I, J=1,\ldots,6$. The gamma matrices $\Gamma^I_{AB}$ satisfy the antisymmetry and reality conditions
\begin{eqnarray}
\Gamma^{IAB}=-\Gamma^{IBA},\quad
\Gamma^{I*}_{AB}=\Gamma^{IAB}=\frac{1}{2}\vp^{ABCD}\Gamma^I_{CD}.
\end{eqnarray}
The $SU(4)$ generators are defined as
\begin{equation}\label{su4g}
\Sigma^{IJ}_A{}^B=\frac{1}{4}(\Gamma^I_{AC}\Gamma^{JCB}-\Gamma^J_{AC}\Gamma^{ICB}).
\end{equation}

\section{A review of the $\CN=5$ theory }\label{SecN5Action}

The $\CN=5$ action is given by \cite{Hosomichi:2008jb,ChenWu3}
\begin{eqnarray}\label{5lagran}\nonumber
{\cal L}&=&\frac{1}{2}(-D_\mu\bar Z^A_aD^\mu
Z^a_A+i\bar\psi^A_a\gamma_\mu
D^\mu\psi^a_A)-\frac{i}{2}\omega^{AB}\omega^{CD}k_{mn}(\cj^m_{AC}\cj^n_{BD}-
2\cj^m_{AC}\cj^n_{DB})\nonumber\\
&&+\frac{1}{2}\epsilon^{\mu\nu\lambda}(k_{mn}A_\mu^m\partial_\nu
A_\lambda^n+\frac{1}{3}C_{mnp}A_\mu^mA_\nu^nA_\lambda^p)\\
&&+\frac{1}{30}C_{mnp}\mu^{mA}{}_B\mu^{nB}{}_C\mu^{pC}{}_A
+\frac{3}{20}k_{mp}k_{ns}(\t^m\t^n)_{ab}Z^{Aa}Z^b_A\mu^{pB}{}_C\mu^{sC}{}_B.\nonumber
\end{eqnarray}
Here $A=1,\ldots,4$ is a fundamental index of the $USp(4)$ R-symmetry group, and  $\omega_{AB}$ is the antisymmetric form of $USp(4)$, satisfying $\omega_{AB}\omega^{BC}=\d^C_A$. The gauge group index $a$ runs from $1$ to $2L$, and the algebra of the gauge group must be chosen as the bosonic subalgebra
of the superalgebra,
\e\label{SLie} [M^m, M^n]=C^{mn}{}_sM^s,\quad
[M^m, Q_a]=-\t^m_{ab}\omega^{bc}Q_c,\quad
\{Q_a,Q_b\}=\t^m_{ab}k_{mn}M^n,\ee
where $k_{mn}=k_{nm}$ and $\omega_{ab}=-\omega_{ba}$ are invariant forms on the superalgebra. The inverse of $\omega_{ab}$ is denoted
as $\omega^{bc}$, satisfying $\omega_{ab}\omega^{bc}=\d_a^c$; and we use $\omega$ to lower or raise indices (for instance, $\t^m_{ab}=\omega_{ac}\t^{mc}{}_b$). We also use $k_{mn}$ to lower indices; for instance, $C_{mnp}=k_{mq}k_{ns}C^{qs}{}_p$.
Following Ref. \cite{GaWi}, in (\ref{5lagran}), we have defined the
``momentum map" and ``current" operators as
\e\label{n5maps}
\mu^m_{AB}\equiv \t^m_{ab}Z^a_AZ^b_B, \quad
\mathcal{J}^m_{AB}\equiv\t^m_{ab}Z^a_A\p^b_B.\ee
The matter fields satisfy the reality conditions
\begin{eqnarray}\label{RealCondi}
\bar Z^A_a=Z_{A}^{*a}=\omega^{AB}\omega_{ab}Z^b_B ,\quad
\bar\psi^A_a=\psi_{A}^{*a}=\omega^{AB}\omega_{ab}\psi^b_B.
\end{eqnarray}

The $\CN=5$ super poincare symmetry transformations are given by
\begin{eqnarray}\label{n5susy}\nonumber
\delta_\ep Z^a_A&=&i\epsilon_A{}^{B}\psi^a_{B},\nonumber\\
\delta_\ep\psi^a_{A}&=&-\gamma^{\mu}D_\mu Z^a_B\epsilon_A{}^B
-\frac{1}{3}k_{mn}\t^{ma}{}_{b}\omega^{BC}Z^b_B\mu^n_{CD}\epsilon_A{}^D
+\frac{2}{3}k_{mn}\t^{ma}{}_{b}\omega^{BD}Z^b_C\mu^n_{DA}\epsilon_B{}^C,
\nonumber\\
\delta_\ep A^m_\mu&=& i\epsilon^{AB}\gamma_\mu \cj^m_{AB}.
\end{eqnarray}
The set of antisymmetric parameters $\ep_A{}^B=\ep^I\Gamma^I_A{}^B$ ($I=1,\ldots, 5$) are required to obey the traceless conditions and the reality conditions
\begin{eqnarray}\label{SusyPara5}\nonumber
\omega_{AB}\epsilon^{AB}=0 ,\quad
\epsilon^{*}_{AB}=\omega^{AC}\omega^{BD}\epsilon_{CD}.
\end{eqnarray}
(The $SO(5)$ gamma matrices $\Gamma^I_A{}^B$ are defined in Appendix \ref{secSO5}.)

The reason for choosing the bosonic subalgebra of (\ref{SLie}) as the
algebra of gauge symmetry is the following: In constructing the $\CN=4$ theory of
Gaiotto and Witten (GW) \cite{GaWi},
GW showed that in order to promote the supersymmetry
from $\CN=1$ to $\CN=4$, the representation matrices of the algebra
of gauge symmetry must satisfy the ``fundamental identity"
\e\label{EFI}
k_{mn}\t^m_{(ab}\t^n_{c)d}=0.
\ee
GW \cite{GaWi} solved the above equation by converting it into the $QQQ$ Jacobi identity of the superalgebra (\ref{SLie}). Eq. (\ref{EFI}) must also hold in the $\CN=5$ theory, since the $\CN=5$ theory can be derived as a special case of the $\CN=4$ theory \cite{Hosomichi:2008jb}. The superalgebra (\ref{SLie}) can be specified as one of the superalgebras
\begin{equation}\label{listsa}
U(M|N),\quad OSp(M|2N),\quad  OSp(2|2N),\quad F(4),\quad G(3),\quad
D(2|1;\a),
\end{equation}
where $\a$ is a continuous parameter. 

\section{Derivation of the $\CN=5$ Super Poincare Currents}\label{SecDN5C}
In this Appendix, we will construct the $\CN=5$ super poincare currents by using the Noether method. According to our discussion in Section \ref{SecN51}, we expect that under the super poincare transformations (\ref{n5susy}), the variation of the $\CN=5$ action (\ref{5lagran}) takes the form
\e\label{EdS2}
\d_\ep S=\int d^3x(-j^I_\mu)\partial^\mu\ep^I,
\ee
where the set of parameters $\ep^I$ $(I=1,\ldots,5)$ are allowed to depend on the spacetime coordinates $x^\mu$. Let us begin by calculating the variations of the kinematic terms of matter fields. They are given by
\e\label{dZ}
\d_\ep(-\frac{1}{2}D_\mu\bar Z^A_aD^\mu
Z^a_A)&=&-\partial^\mu(D_\mu\bar Z^A_a\d_\ep Z^a_A)\nonumber\\
&&+i\ep_A{}^B\psi^a_BD^2\bar Z^A_a\nonumber\\
&&+i\ep_A{}^B\g_\mu\cj^{mA}{}_BD_\mu
\bar Z^C_a\t_m{}^a{}_bZ^b_C\ee
and
\e\label{dpsi}
&&\d_\ep (\frac{i}{2}\bar\psi^A_a\gamma_\mu
D^\mu\psi^a_A)\nonumber\\
&=&-\frac{i}{2}\partial_\mu(\bar\psi^A_a\g^\mu\d_\ep\psi^a_A)\nonumber\\
&&+i\bar\psi^A_a\g^\mu(\d\psi)^{Ia}_A\partial_\mu\ep^I\nonumber\\
&&-i\ep_{A}{}^B\bar\psi^A_aD^2Z^a_B+\ep_A{}^B\g^{\mu\nu}\cj_{mB}{}^AF^m_{\mu\nu}\nonumber\\
&&+\frac{i}{3}(\ep_A{}^B\g^\mu\bar\psi^A_a)\t^{ma}{}_bD_\mu(Z^b_C\mu_m^C{}_B)-\frac{2i}{3}(\ep_A{}^B\g^\mu\bar\psi^C_a)\t^{ma}{}_bD_\mu(Z^b_B\mu_m^A{}_C)\nonumber\\
&&-(\ep_A{}^B\t_m{}^a{}_b\psi^b_C)(\bar\psi^C_a\cj^{mA}{}_B),
\ee
where $\g^{\mu\nu}$ is defined by (\ref{gammamunu1}), and $(\d\psi)^{Ia}_A$ is defined via the $\d_\ep\psi^{a}_A= (\d\psi)^{Ia}_A\ep^I$, i.e.
\begin{equation}\label{svofp2}
(\d\psi)^{Ia}_A\equiv-\gamma^{\mu}D_\mu Z^a_B\Gamma^I_A{}^B
-\frac{1}{3}k_{mn}\t^{ma}{}_{b}\omega^{BC}Z^b_B\mu^n_{CD}\Gamma^I_A{}^D
+\frac{2}{3}k_{mn}\t^{ma}{}_{b}\omega^{BD}Z^b_C\mu^n_{DA}\Gamma^I_B{}^C.
\end{equation}

The super poincare variation of the Chern-Simons term is given by
\e\label{dcs}
\d_\ep\cal{L}_{{\rm CS}}&=&\frac{1}{2}\ep^{\mu\nu\lambda}k_{mn}\partial_{\nu}(A^m_\mu \d_\ep A^n_\lambda)\nonumber\\
&&+\ep_A{}^B\g^{\mu\nu}k_{mn}\cj^{mA}{}_BF^n_{\mu\nu}.
\ee

Writing the Yukawa terms in (\ref{5lagran}) as
\e\label{yterms}
\CL_{{\rm Y}}=\frac{i}{2}\cj_m^A{}_C\cj^{m}_A{}^C
-i\cj_m^A{}_C\cj^{mC}{}_A,
\ee
then
\e\label{dy}
\d_\ep\CL_{{\rm Y}}
&=&\nonumber\t^m_{ab}(\ep_A{}^B\psi^a_B)(\cj^{AC}_m\psi^b_C -2 \psi^b_C\cj^{CA}_m) \\ &&+i\t^m_{ab}(\ep_A{}^B\g^\mu\cj_{mC}{}^A)Z^{Ca}D_\mu Z^b_B-2i\t^m_{ab}(\ep_A{}^B\g^\mu\cj^{mA}{}_C)Z^{Ca}D_\mu Z^b_B\nonumber
 \\ &&+\frac{i}{3}(\t_m\t_n)_{ab}\ep_{AB}\big(\cj^{mCB}Z^a_CZ^b_D\mu^{nDA}
-2\cj^{mBC}Z^a_CZ^b_D\mu^{nDA}\nonumber\\ &&\quad\quad\quad\quad\quad\quad\quad\quad
+2\cj^{mCD}Z^a_CZ^{Bb}\mu^{n}{}_D{}^A-4\cj^{mCD}Z^a_DZ^{Bb}\mu^{n}{}_C{}^A\big),
\ee
with
\e\label{tautau}
(\t_m\t_n)_{ab}=\t_{mac}\t_n{}^c{}_b=\frac{1}{2}(\{\t_m,\t_n\}_{ab}+[\t_m,\t_n]_{ab})
=(\t_m\t_n)_{[ab]}+\frac{1}{2}C_{mnp}\t^p_{ab}.
\ee

Finally, the super variation of the potential reads
\e\label{dpot}
\d_\ep\CL_{{\rm pot}}&=&\frac{i}{10}C_{mnp}\ep_{AB}\cj^{mCB}\mu^{nA}{}_D\mu^{pD}{}_C\nonumber\\
&&-\frac{3i}{5}(\t_m\t_n)_{[ab]}\ep_{AB}\cj^{mCB}(Z^{a}_DZ^{Ab}\mu^{nD}{}_C+Z^{Da}Z^b_D\mu^{nA}{}_C).
\ee
Dropping the total derivative terms, and adding everything together, we obtain
\begin{equation}
\d_\ep S=\int d^3x\big[i\bar\psi^A_a\g^\mu(\d\psi)^{Ia}_A\partial_\mu\ep^I+(\psi^3Z\;{\rm terms})+(\psi (DZ)Z^2\;{\rm terms})+(\psi Z^5\;{\rm terms})\big].
\end{equation}
Here we use ``$\psi^3Z$ terms" to represent all terms that are third order in $\psi$ and first order
in $Z$, and ``$\psi Z^5$ terms" has a similar meaning. And we use ``$\psi (DZ)Z^2$ terms"
to present all terms that are proportional to $\psi (DZ)Z^2$, with $D$
the covariant derivative.

We will prove that
\begin{equation}
0=(\psi^3Z\;{\rm terms})=(\psi (DZ)Z^2\;{\rm terms})=(\psi Z^5\;{\rm terms}).
\end{equation}
Let us first prove that all the $\psi^3Z$ terms cancel;
they are given by (see the last line of (\ref{dpsi}) and the first line of (\ref{dy}))
\e\label{psi3z1}
-(\ep_A{}^B\t_m{}^a{}_b\psi^b_C)(\bar\psi^C_a\cj^{mA}{}_B)+\t^m_{ab}(\ep_A{}^B\psi^a_B)
(\cj^{AC}_m\psi^b_C -2 \psi^b_C\cj^{CA}_m).
\ee
To prove that (\ref{psi3z1}) vanishes, we need the following two key identities
\e
\label{totantiR}
&&\omega_{[AB}\omega_{CD}\omega_{EF]}=0,\\
&&\ep_{[\a\b}\ep_{\g\d]}=0.
\label{totantiS}
\ee
Using the $USp(4)$ identity (\ref{totanti}), one can show that (\ref{totantiR}) is equivalent to the well-known identity:
\e
\vp_{GABC}\vp^{GDEF}=3!\d^{[A}_D\d^B_E\d^{C]}_F.
\ee
Also, notice that (\ref{totantiS}) is equivalent to
\e\label{totantiS2}
(\psi_1\psi_2)\psi_3+(\psi_2\psi_3)\psi_1+(\psi_3\psi_1)\psi_2=0,
\ee
where $\psi_1$, $\psi_2$, and $\psi_3$ are arbitrary spinors.

Multiplying both sides of (\ref{totantiR}) by the parameters $\ep^{EF}$, one obtains the identity
\e\label{totantiR2}
\omega_{[AB}\ep_{CD]}=0.
\ee
Multiplying both sides of (\ref{totantiR2}) by $\t^m_{ab}\psi^{Ba}(\psi^{Db}\cj^{CA}_m)$,
we have
\e\label{psi3z12}\nonumber
0&=&-\t^m_{ab}(\ep_{AB}\psi^{Ca})(\psi^{Bb}\cj_m^{A}{}_C-\psi^b_C\cj_m^{AB}-\psi^{Bb}\cj_{mC}{}^A)\\
&&+\t^m_{ab}(\ep_{AB}\psi^{Ba})(\psi^{Ab}\cj_{mC}{}^C+\psi^b_C\cj_m^{CA}+\psi^{Cb}\cj_m^A{}_C).
\ee
Using (\ref{totantiS2}), the last term of the first line of (\ref{psi3z12}) can be converted into:
\begin{equation}\label{psi3z13}
\t^m_{ab}(\ep_{AB}\psi^{Ca})\psi^{Bb}\cj_{mC}{}^A=
-\t^m_{ab}(\ep_{AB}\psi^{Bb})(\psi^{Ca}\cj_{mC}{}^A)-\t^m_{ab}(\ep_{AB}\cj_{mC}{}^A)(\psi^{Bb}\psi^{Ca})
\end{equation}
Using the fundamental identity $k_{mn}\t^m_{(ab}\t^n_{c)d}=0$ (see (\ref{EFI})), the second term of the RHS of (\ref{psi3z13}) can be decomposed into two terms:
\begin{equation}\label{psi3z14}
-\t^m_{ab}(\ep_{AB}\cj_{mC}{}^A)(\psi^{Bb}\psi^{Ca})=-\t^m_{ab}(\ep_{AB}\psi^{Ba})(\psi^{Ab}\cj_{mC}{}^C
+\psi^{Cb}\cj_{mC}{}^A).
\end{equation}
Combining (\ref{psi3z12})$-$(\ref{psi3z14}) gives
\begin{equation}\label{psi3z15}
\t^m_{ab}[(\ep_{AB}\psi^{Ca})(\psi^b_C\cj^{AB}_m-\psi^{Bb}\cj_{m}^A{}_C)+(\ep_{AB}\psi^{Bb})
(3\psi^a_C\cj^{CA}_m+\psi^{Cb}\cj_m^A{}_C)]=0.
\end{equation}
Notice that the second term of (\ref{psi3z15}) has the two different expressions:
\e\label{psi3z16}
-\t^m_{ab}(\ep_{AB}\psi^{Ca})(\psi^{Bb}\cj_{m}^A{}_C)&=&\t^m_{ab}[(\ep_{AB}\cj^{AC}_m)(\psi^{Ba}\psi^b_
C)+(\ep_{AB}\psi^{Ca})(\psi^b_C\cj_m^{AB})]\\
&=&-\t^m_{ab}[(\ep_{AB}\cj^{AC}_m)(\psi^{Ba}\psi^b_
C)-(\ep_{AB}\psi^{Ba})(\psi^b_C\cj_m^{AC})].\nonumber
\ee
We have used (\ref{EFI}) and (\ref{totantiS2}) to derive the first line and  the second line, respectively. Taking the average of the two lines of (\ref{psi3z16}) gives
\begin{equation}\label{psi3z17}
-\t^m_{ab}(\ep_{AB}\psi^{Ca})(\psi^{Bb}\cj_{m}^A{}_C)=\frac{1}{2}
\t^m_{ab}[(\ep_{AB}\psi^{Ca})(\psi^b_C\cj_m^{AB})
+(\ep_{AB}\psi^{Ba})(\psi^b_C\cj_m^{AC})].
\end{equation}
Plugging (\ref{psi3z17}) into (\ref{psi3z15}), one obtains
\begin{equation}
\frac{3}{2}\t^m_{ab}[(\ep_{AB}\psi^{Bb})(2\psi^a_C\cj^{CA}_m+\psi^{Cb}\cj_m^A{}_C)+(\ep_{AB}\psi^{Ca})(\psi^b_C\cj^{AB}_m)]=0.
\end{equation}
Notice that the left hand side is proportional to (\ref{psi3z1}), so (\ref{psi3z1}) does \emph{vanish}, i.e. all $\psi^3Z$ terms cancel.

Let us now try to prove that the $\psi (DZ)Z^2$ terms cancel; they are given by the last line of (\ref{dZ}), the fourth line of (\ref{dpsi}), and the second line of (\ref{dy}):
\e\label{psidz3}
&&+i\ep_A{}^B\g_\mu\cj^{mA}{}_BD_\mu
\bar Z^C_a\t_m{}^a{}_bZ^b_C\nonumber\\
&&+\frac{i}{3}(\ep_A{}^B\g^\mu\bar\psi^A_a)\t^{ma}{}_bD_\mu(Z^b_C\mu_m^C{}_B)-\frac{2i}{3}(\ep_A{}^B\g^\mu\bar\psi^C_a)\t^{ma}{}_bD_\mu(Z^b_B\mu_m^A{}_C)\nonumber\\
 &&+i\t^m_{ab}(\ep_A{}^B\g^\mu\cj_{mC}{}^A)Z^{Ca}D_\mu Z^b_B-2i\t^m_{ab}(\ep_A{}^B\g^\mu\cj^{mA}{}_C)Z^{Ca}D_\mu Z^b_B.
\ee
Using the fundamental identity $k_{mn}\t^m_{(ab}\t^n_{c)d}=0$, (\ref{psidz3}) can be recast as
\e
\frac{2i}{3}\big(\omega_{[CD}\ep_{AB]}\g^\mu\cj^{AB}_m\big)\t^m_{ab}(D_\mu Z^{Ca})Z^{Db},
\ee
which \emph{vanishes} on account of the identity (\ref{totantiR2}).

Finally, we need to take care of the $\psi Z^5$ terms; they are given by (\ref{dpot}) and the last two lines of (\ref{dy}): (We denote them as $\mathcal{O}(Z^5)$.)
\e\label{psiz5}
&&\mathcal{O}(Z^5)\nonumber\\&\equiv&\frac{i}{10}C_{mnp}\ep_{AB}\cj^{mCB}\mu^{nA}{}_D\mu^{pD}{}_C\nonumber\\
&&-\frac{3i}{5}(\t_m\t_n)_{[ab]}\ep_{AB}\cj^{mCB}(Z^{a}_DZ^{Ab}\mu^{nD}{}_C
+Z^{Da}Z^b_D\mu^{nA}{}_C).\nonumber\\
&&+\frac{i}{3}[(\t_m\t_n)_{[ab]}+\frac{1}{2}C_{mnp}\t^p_{ab}]\ep_{AB}\big(\cj^{mCB}Z^a_CZ^b_D\mu^{nDA}
-2\cj^{mBC}Z^a_CZ^b_D\mu^{nDA}\nonumber\\ &&\quad\quad\quad\quad\quad\quad\quad\quad
+2\cj^{mCD}Z^a_CZ^{Bb}\mu^{n}{}_D{}^A-4\cj^{mCD}Z^a_DZ^{Bb}\mu^{n}{}_C{}^A\big).
\ee
Notice that some terms are proportional to $(\t_m\t_n)_{[ab]}$, while the rest are proportional to $C_{mnp}\t^p_{ab}$. However, using the identity \footnote{This identity can be proved as follows:
\e\nonumber
C_{mnp}\t^n_{cd}\t^p_{ab}&=&[\t_m, \t_n]_{ab}\t^n_{cd}\nonumber\\
&=&\t_{mae}\t_n{}^e{}_b\t^n_{cd}-\t_{nae}\t_m{}^e{}_b\t^n_{cd}\nonumber\\
&=&-(\t_m\t_n)_{ac}\t^n_{bd}-(\t_m\t_n)_{ad}\t^n_{cb}
+(\t_n\t_m)_{db}\t^n_{ac}+(\t_n\t_m)_{cb}\t^n_{ad}.\nonumber
\ee
In the third line, we have used the fundamental identity $k_{mn}\t^m_{(ab}\t^n_{c)d}=0$. Substituting $(\t_m\t_n)_{ab}=(\t_m\t_n)_{[ab]}+\frac{1}{2}C_{mnp}\t^p_{ab}$ (see (\ref{tautau})) into the last line, and simplifying the expression, one obtains the identity (\ref{newid}).}
\begin{equation}\label{newid}
C_{mnp}\t^n_{cd}\t^p_{ab}=(\t_m\t_n)_{[ca]}\t^n_{bd}+(\t_m\t_n)_{[da]}\t^n_{cb}+
(\t_m\t_n)_{[db]}\t^n_{ac}+(\t_m\t_n)_{[cb]}\t^n_{ad},
\end{equation}
we are able to recast (\ref{psiz5}) so that every term is proportional to $(\t_m\t_n)_{[ab]}$:
\e\label{psiz52}
&&\mathcal{O}(Z^5)\nonumber\\&=&\frac{2i}{15}(\ep_{AB}\cj^{mCB})(\t_m\t_n)_{[ab]}(3Z^a_CZ^b_D\mu^{nDA}+4Z^a_DZ^{Db}\mu^{nA}{}_C
+5Z^{Aa}Z^b_D\mu^{nD}{}_C)\nonumber\\&&
+\frac{i}{3}(\ep_{AB}\cj^{mBC})(\t_m\t_n)_{[ab]}(-3Z^a_CZ^b_D\mu^{nDA}+Z^a_DZ^{Db}\mu^{nA}{}_C
-Z^{Aa}Z^b_D\mu^{nD}{}_C)
\nonumber\\&&
+\frac{i}{3}(\ep_{AB}\cj^{mCD})(\t_m\t_n)_{[ab]}(Z^{Aa}Z^b_C\mu^{nB}{}_D
-5Z^{Aa}Z^{b}_D\mu^{nB}{}_{C}
-Z^{Aa}Z^{Bb}\mu^{n}_{CD}).
\ee
We now have to prove that (\ref{psiz52}) vanishes. To simplify the task, let us introduce some shorthands
\e\label{shorthands}
X_1&\equiv&(\ep_{AB}\cj^{mCB})(\t_m\t_n)_{[ab]}Z^a_CZ^b_D\mu^{nDA},\nonumber\\
X_2&\equiv& -(\ep_{AB}\cj^{mCB})(\t_m\t_n)_{[ab]}Z^a_DZ^{Db}\mu^{nA}{}_C,\nonumber\\
X_3&\equiv& (\ep_{AB}\cj^{mCB})(\t_m\t_n)_{[ab]}Z^{Aa}Z^b_D\mu^{nD}{}_C,\nonumber\\
Y_1&\equiv& (\ep_{AB}\cj^m_C{}^C)(\t_m\t_n)_{[ab]}Z^a_DZ^{Bb}\mu^{nDA},\nonumber\\
Y_2&\equiv& (\ep_{AB}\cj^{mBC})(\t_m\t_n)_{[ab]}Z^{Aa}Z^{b}_D\mu^{nD}{}_C,
\nonumber\\
Y_3&\equiv& (\ep_{AB}\cj^{mBC})(\t_m\t_n)_{[ab]}Z^{a}_DZ^{Db}\mu^{nA}{}_C,
\nonumber\\
Y_4&\equiv& (\ep_{AB}\cj^{mBC})(\t_m\t_n)_{[ab]}Z^a_CZ^{b}_D\mu^{nDA},
\nonumber\\
Y_5&\equiv& (\ep_{AB}\cj^{mCD})(\t_m\t_n)_{[ab]}Z^{Ba}Z^{b}_D\mu^{nA}{}_C,
\nonumber\\
Y_6&\equiv& (\ep_{AB}\cj^{mCD})(\t_m\t_n)_{[ab]}Z^{Aa}Z^{Bb}\mu^{n}_{CD},
\nonumber\\
Y_7&\equiv& (\ep_{AB}\cj^{mCD})(\t_m\t_n)_{[ab]}Z^{Aa}Z^{b}_C\mu^{nB}{}_{D}.
\ee
In terms of the shorthands, (\ref{psiz52}) reads
\e\label{psiz53}
\mathcal{O}(Z^5)=\frac{2i}{15}(3X_1-4X_2+5X_3)+\frac{i}{3}(-3Y_4+Y_3-Y_2)+\frac{i}{3}(Y_7+5Y_5-Y_6).
\ee
To prove that (\ref{psiz53}) vanishes, we need to establish some connections between the quantities defined in (\ref{shorthands}). Taking account of the identity (\ref{totantiR}) and (\ref{totantiR2}), we have
\e
0&=&\omega_{[BE}\omega_{CF}\omega_{DG]}(\ep_{A}{}^E\cj^{mCB})
(\t_m\t_n)_{[ab]}Z^{Fa}Z^{Gb}\mu^{nDA},\\
0&=&(\omega_{[CD}\ep_{AB]}\cj^{mDB})
(\t_m\t_n)_{[ab]}Z^{Ca}Z^{b}_E\mu^{nEA},\\
0&=&(\omega_{[CD}\ep_{AB]}\cj^{mCE})
(\t_m\t_n)_{[ab]}Z^{Aa}Z^{Bb}\mu^{nD}{}_E,
\ee
which can be simplified to give
\e\label{c36}
0&=&(2X_1-X_2)+(-2Y_1-Y_3-2Y_4+2Y_5+2Y_7),\\
0&=&(X_1-X_3)+(-Y_1+Y_2-Y_4),\label{c44}\\
0&=&2Y_2+Y_3+Y_6-2Y_7,\label{c47}
\ee
respectively. On the other hand,
using (\ref{EFI}), (\ref{tautau}), and (\ref{newid}), we are able to decompose $Y_1$ as follows
\e\label{y1}
2Y_1=Y_2-3Y_3+3Y_4-3Y_5+3Y_6+Y_7.
\ee
Combining (\ref{c36}), (\ref{c44}), and (\ref{psiz53}), one obtains
\e\label{psiz54}
\mathcal{O}(Z^5)=\frac{i}{15}(6Y_1+5Y_2+13Y_3-9Y_4+9Y_5-5Y_6-11Y_7).
\ee
Plugging (\ref{y1}) into the above equation gives
\e
\mathcal{O}(Z^5)=\frac{4i}{15}(2Y_2+Y_3+Y_6-2Y_7)=0,
\ee
which is nothing but the identity (\ref{c47}). This completes the proof that all $\psi Z^5$ terms cancel.

In summary, we have
\begin{equation}
\d_\ep S=\int d^3x(i\bar\psi^A_a\g^\mu(\d\psi)^{Ia}_A\partial_\mu\ep^I).
\end{equation}
Hence the $\CN=5$ super poincare currents are given by
\e
j^I_\mu=-i\bar\psi^A_a\g_\mu(\d\psi)^{Ia}_A.
\ee


\begin{thebibliography}{99}

\bibitem{Hosomichi:2008jb}
  K.~Hosomichi, K.~M.~Lee, S.~Lee, S.~Lee and J.~Park,
 ``N=5,6 Superconformal Chern-Simons Theories and M2-branes on Orbifolds,''
 J. High Energy Phys. 09 (2008) 002,
  arXiv:0806.4977 [hep-th].

\bibitem{Chen2}
Fa-Min Chen, ``Symplectic  Three-Algebra Unifying ${\cal N}=5,6$
Superconformal Chern-Simons-Matter Theories," J. High Energy Phys. 08 (2010) 077, arXiv:0908.2618 [hep-th].



\bibitem{ChenWu3}
Fa-Min Chen, Yong-Shi Wu, ``Superspace Formulation in a
Three-Algebra Approach to $D=3, {\cal N}=4,5$ Superconformal
Chern-Simons Matter Theories,'' Phys. Rev. D,  Volume 82, Issue 10, arXiv: 1007.5157 [hep-th].

\bibitem{Schwarz}
  M.~Bandres, A.~Lipstein and J.~Schwarz
 ``Studies of the ABJM Theory in Formulation with Manifest $SU(4)$ R-Symmetry,'' J. High Energy Phys. 09 (2008) 027,
  arXiv:0807.0880 [hep-th].

\bibitem{GaWi}
  D.~Gaiotto and E.~Witten,
   ``Janus Configurations, Chern-Simons Couplings, And The Theta-Angle in N=4
  Super Yang-Mills Theory,''
  J. High Energy Phys. 06 (2010) 097,
  arXiv:0804.2907 [hep-th].

\bibitem{Aharony:2008gk}
  O.~Aharony, O.~Bergman and D.~L.~Jafferis,
  ``Fractional M2-branes,''
  J. High Energy Phys. 11 (2008) 043,
  arXiv:0807.4924 [hep-th].

\bibitem{Chen:pku1} Fa-Min Chen, ``Superalgebra Realization of the 3-algebras in N=6, 8 Chern-Simons-matter Theories," J. Math. Phys. 53, 012301 (2012); arXiv:1012.0904. [hep-th].

\bibitem{ABJM}
O.~Aharony, O.~Bergman, D.~L.~Jafferis and J.~Maldacena,
  ``N=6 superconformal Chern-Simons-matter theories, M2-branes and their
  gravity duals,''
  J. High Energy Phys. 10 (2008) 091,
  arXiv:0806.1218 [hep-th].

\bibitem{HosomichiJD}
  K.~Hosomichi, K.~M.~Lee, S.~Lee, S.~Lee and J.~Park,
  ``N=4 Superconformal Chern-Simons Theories with Hyper and Twisted Hyper
  Multiplets,'' J. High Energy Phys. 07 (2008) 091,
  arXiv:0805.3662 [hep-th].

\bibitem{Palmkvist1}
Jakob Palmkvist, ``Unifying $\CN=5$ and $\CN=6$", JHEP 1105: 088, 2011, arXiv:1103.4860 [hep-th].

\bibitem{chen5}
Fa-Min Chen, ``Covariantly Constant Curvature Tensors and D=3, N=4, 5, 8 Chern-Simons Matter Theories," Phys. Rev. D 85, 065017 (2012), arXiv:1109.3865 [hep-th].

\bibitem{Palmkvist2}
Sung-Soo Kim, Jakob Palmkvist, ``$\CN=5$ three-algebras and 5-graded Lie superalgebras", J. Math. Phys. 52: 083502, 2011, arXiv:1010.1457 [hep-th].

\bibitem{Bagger}
  J.~Bagger and N.~Lambert,
  ``Modeling multiple M2's,''
  Phys.\ Rev.\  D {\bf 75}, 045020 (2007),
  arXiv:hep-th/0611108;
``Gauge Symmetry and Supersymmetry of Multiple M2-Branes,''
  Phys.\ Rev.\  D {\bf 77}, 065008 (2008),
  arXiv:0711.0955 [hep-th];
  J.~Bagger and N.~Lambert,
  ``Comments On Multiple M2-branes,''
  JHEP {\bf 0802}, 105 (2008),
  arXiv:0712.3738 [hep-th].

\bibitem{Gustavsson}
  A.~Gustavsson,
  ``Algebraic structures on parallel M2-branes,'' Nucl. Phys. 811B, 66 (2009),
  arXiv:0709.1260 [hep-th];
``Selfdual strings and loop space Nahm equations,''
  JHEP {\bf 0804}, 083 (2008),
  arXiv:0802.3456 [hep-th].

\bibitem{Schwarz0}
  M.~Bandres, A.~Lipstein and J.~Schwarz
 ``$\CN = 8$ Superconformal Chern--Simons Theories,'' JHEP 0805: 025, 2008,
  arXiv:0803.3242 [hep-th].

\bibitem{ChenWu1}
Fa-Min Chen, Yong-Shi Wu, ``Symplectic Three-Algebra and ${\cal
N}=6, Sp(2N)\times U(1)$ Superconformal Chern-Simons-Matter Theory
," European Physics Journal C (2010), 69: 305-314, arXiv:0902.3454 [hep-th].



\end{thebibliography}
\end{document}